\documentclass[journal,transmag]{IEEEtran}
\IEEEoverridecommandlockouts
\usepackage[T1]{fontenc}
\usepackage[utf8]{inputenc}   % keep once
\usepackage{lmodern}          % scalable Latin Modern fonts
\usepackage{microtype}        % typographical refinements

\usepackage{amsmath,amssymb,amsfonts}
\usepackage{amsthm}
\newtheoremstyle{mystyle}% name
  {}{}% space above / below
  {\itshape}% body font
  {}% indent
  {\bfseries}% head font
  {.}% punctuation
  { }% space after head
  {\thmname{#1}\thmnumber{ #2}\thmnote{ (#3)}}
\theoremstyle{mystyle}

\newtheorem{remark}{Remark}

\newcounter{subassumption}[asu]

\makeatletter
\renewcommand{\p@subassumption}{\theasu}
\makeatother

\usepackage{graphicx}
\usepackage{epstopdf}    % convert EPS to PDF (optional)
\usepackage{booktabs}
\usepackage{graphicx} 
\usepackage{mwe} 
\usepackage{array}
\usepackage{tabularx}
\usepackage{multirow}
\usepackage{wrapfig}     % optional: wrapped figures
\usepackage{float}       % provides [H]

\usepackage{algorithm}
\usepackage{algpseudocode}  % provides \begin{algorithmic}[1]

\usepackage{cite}                % compress/sort numeric citations
\PassOptionsToPackage{hyphens}{url}
\usepackage[hyphens,spaces,obeyspaces]{url}
\usepackage[english]{babel}

\usepackage{xcolor}       % color support (preferred over 'color')
\usepackage{ragged2e}
\usepackage{verbatim}
\usepackage{csquotes}
\usepackage{textcomp}     % extra symbols
\usepackage{lineno,nohyperref}  % linenumbers without hyperref
\usepackage[bottom]{footmisc}   % footnotes at bottom
\usepackage{authblk}      % if you use \affil (optional)
\usepackage{blindtext}    % for testing; remove in final
\usepackage{xspace}       % safe use in custom macros

\newcommand{\vect}[1]{\boldsymbol{#1}}
\newcommand*{\affaddr}[1]{#1}
\newcommand*{\affmark}[1][*]{\textsuperscript{#1}}
\newcommand*{\email}[1]{#1}

\def\BibTeX{{\rm B\kern-.05em{\sc i\kern-.025em b}\kern-.08em
    T\kern-.1667em\lower.7ex\hbox{E}\kern-.125emX}}

\begin{document}
 	\title{%Satellite-Assisted Intent-Aware Integrated Access and Backhaul for Connecting Rural Areas
    %AI-Driven Intent-Aware Satellite–Terrestrial Integrated Access and Backhaul for  Rural Areas
 Large Language Model Assisted Intent-Based Satellite-Integrated Access and Backhaul FWA for Rural Areas
    \\
		\thanks{This work was supported by NSERC (under project ALLRP 566589-21) and  InnovÉÉ (INNOV-R program) through the partnership with Ericsson and ECCC.
		We thank the Ericsson's Montréal
		GAIA team for their constructive and helpful comments, which have
		significantly improved the quality and clarity of this manuscript (corresponding author: Anselme Ndikumana).}}
		\author{%
	Anselme Ndikumana\affmark[1], Kim Khoa Nguyen \affmark[1], Adel Larabi\affmark[2], and Mohamed Cheriet\affmark[1]\\
	\affaddr{\affmark[1]Synchromedia Lab, École de Technologie Supérieure, Université du Québec, QC, Canada\\  \email{\{anselme.ndikumana, kim-khoa.nguyen; Mohamed.Cheriet\}}@etsmtl.ca}\\
	\affaddr{\affmark[2] GAIA  Montreal, Ericsson Canada}\\
	\email{\{adel.larabi\}@ericsson.com}\\
	}
\maketitle  
\begin{abstract}
Rural areas exhibit low population density and highly variable connectivity needs shaped by both household usage and field operations such as planting, harvesting, and mining. These field activities often occur in isolated locations requiring temporary connectivity, whereas rural households depend on fixed broadband. During intensive outdoor activities, household fixed networks may remain underutilized, resulting in inefficient resource use and unnecessary energy consumption. The coexistence of residential and field-based communication demands creates substantial spatial and temporal fluctuations that the current rural network cannot effectively adapt to. Limited visibility into user mobility, activity patterns, and intent makes it difficult for operators to coordinate temporary and fixed networks. To address these underexplored challenges, we propose an AI-driven Intent-Aware Satellite–Integrated Access and Backhaul (IAB) approach to connect rural areas. In our proposal, a large language model (LLM) translates users' intents into explicit network requirements. Guided by these inferred requirements, we develop a dynamic satellite-IAB-based Fixed Wireless Access (FWA) network approach that jointly optimizes temporary field connectivity and fixed broadband access to maximize energy efficiency while satisfying the data rate requirement. The formulated optimization problem is solved using a two-stage Benders decomposition approach. The simulation results show that our approach significantly reduces energy consumption while maximizing energy efficiency.
\end{abstract}
\begin{IEEEkeywords}
Intent-aware network,  satellite communication, integrated access and backhaul, energy efficiency,  large language model, fixed wireless access
\end{IEEEkeywords}
\section{Introduction}
\subsection{Background and Motivations}
Rural areas play an essential role in national and global economies, being home to major land-based industries such as agriculture, forestry, and natural resource extraction \cite{FAO}. The ongoing digital transformation of these industries increasingly depends on advanced automation and intelligent systems, including autonomous tractors, unmanned aerial vehicles (drones), and collaborative robots. These devices require communication networks to support data exchange, remote monitoring, and real-time control. Unlike urban regions that benefit from dense, permanent communication infrastructure, rural areas are characterized by low population density, wide geographic dispersion, and highly variable traffic demand. Furthermore, field operations, such as planting, harvesting, logging, and mining, are often conducted in isolated locations where connectivity is unavailable. Consequently, field activities require on-demand or temporary networks that can be deployed and activated dynamically when needed.
In contrast, rural households primarily rely on fixed networks, such as Fixed Wireless Access (FWA) networks \cite{Maravedis}, for daily broadband connectivity. In 5G FWA, houses are equipped with Customer Premises Equipment (CPE) with antennas mounted on their roofs, wirelessly connected to fixed cellular base stations \cite{ndikumana2024renewable, ndikumana2026energy, ndikumana2023digital}. However, during outdoor activities, these fixed networks may remain underutilized, resulting in inefficient resource use and unnecessary energy consumption.

The coexistence of residential and field-based communication network needs creates distinct spatial and temporal fluctuations in network demand. Despite recent advances in rural broadband access, the current network infrastructure cannot adapt to dynamic changes in connectivity requirements. Furthermore, network operators typically have limited visibility into population mobility and activity patterns \cite{yao2020understanding}, making it difficult to coordinate between fixed and temporary networks. This gap motivates the development of an intelligent, context-aware approach capable of predicting or interpreting user needs in rural areas to dynamically adjust network configurations and resource allocation. Recent progress in intent-based networking (IBN)\cite{leivadeas2022survey} has opened new possibilities for adaptive network management. In an intent-aware system, users or applications can express high-level operational objectives, referred to as intents, which the network then interprets to provision appropriate connectivity and quality-of-service (QoS) levels automatically. For example, in a rural area,  a farming cooperative could express an intent such as “farm operations from 9:00 AM to 4:00 PM involving drones and robots,” prompting the network to establish temporary high-throughput, low-latency connections in the corresponding area during that time window. Applying such intent awareness in rural networks can significantly improve the coordination between temporary and fixed networks.

Existing rural connectivity solutions, such as FWA, are limited in flexibility, coverage, and adaptability. Integrated Access and Backhaul (IAB) \cite{3GPP38874} has been included in FWA  to extend coverage and capacity. The IAB network consists of an IAB donor and multiple IAB nodes. The IAB donor functions as the central base station and is directly connected to the core network. The remaining base stations, referred to as IAB nodes, establish wireless backhaul links with the IAB donor. In an IAB-based FWA scenario, CPE can connect to either the IAB donor or the IAB nodes via wireless access links \cite{ndikumana2026energy}. Each IAB node is composed of a Distributed Unit (DU) and a Mobile Termination (MT) unit. The DU serves downstream CPEs and subordinate IAB-MTs, while the MT enables the node to operate as a relay by connecting to its parent IAB-DU.
In contrast, the IAB donor integrates both Distributed Unit (DU) and Control Unit (CU) functionalities. However, the IAB donor requires a fiber-optic backhaul to the core network, which is not economically feasible in rural areas. Furthermore, fiber-optic backhaul is often unreliable due to the complex, variable terrain in some rural areas, such as agriculture or mining zones. Hence, there is a need for non-terrestrial network backhauling, such as satellite communication \cite{luo2024leo}, to support terrestrial networks in rural areas to ensure seamless connectivity across diverse operational contexts.

\subsection{Challenging Issues}
Despite significant fixed network progress for rural areas, several key challenges hinder the realization of intent-aware and adaptive connectivity in rural areas:
\begin{itemize}
	\item  FWA \cite{oproiu20185g} provides an affordable solution for broadband connectivity, but is generally limited to one-hop access. Its coverage footprint is inadequate for large, remote areas such as agricultural or forestry sites, and its static nature limits deployment flexibility.
	\item IAB \cite{specification3gpp} extends FWA coverage through multi-hop wireless relays. However, it relies on a fiber-optic backhaul link to the IAB donor, which may be impractical in temporary and remote rural areas.
	\item  Microwave backhaul \cite{ndikumana2025energy} can be effective under clear line-of-sight (LoS) conditions to replace fiber-optic, but its reliability deteriorates when LoS is obstructed by forests, terrain undulations, or vegetation \cite{yaacoub2020efficient}. These limitations reduce its applicability for establishing temporary networks in rural areas.
	\item Satellite communication provides wide-area coverage and inherent mobility \cite{jia2025network}, making it a strong candidate for replacing microwave backhaul to connect the IAB donor to the core network. Nevertheless, integrating satellite links into terrestrial IAB-FWA is unexplored in the existing literature.
\end{itemize}

\subsection{Contributions}
To address the challenges outlined above, this paper proposes an Artificial Intelligence (AI)- driven, intent-aware Satellite–IAB–based FWA approach to connect rural areas. The framework leverages intent-based networking principles to intelligently coordinate the activation, deactivation, and configuration of IAB functionality at terrestrial and satellite nodes based on user activities and connectivity demands in rural areas. The main contributions of this work are summarized as follows:
\begin{itemize}
	\item We propose a mapping approach that converts user intent into network vocabulary and, subsequently, into QoS requirements, using a Large Language Model (LLM) as an AI tool. This enables the automatic establishment of fixed broadband access in residential areas and temporary network coverage in relevant field areas when field activities begin, and the deactivation of unnecessary functions once activities end to minimize energy consumption and release unused resources while maintaining fixed broadband access.
	\item Based on the derived network requirements from intents, we formulate a user admission optimization problem as a 0–1 integer linear program. This approach efficiently resolves conflicts between multiple intents and network requirements, supporting both temporary and fixed network deployments.
	\item  We propose a satellite-assisted IAB network that supports temporary field connectivity alongside FWA. The IAB functions can be dynamically activated or deactivated to maximize energy efficiency (i.e., minimize energy consumption) while meeting delay and data rate requirements.
	\item For network setup and configuration in the satellite-assisted IAB-based  FWA network, we formulate an optimization problem to maximize energy efficiency while satisfying delay and data rate requirements. Then, we propose a two-stage Benders decomposition \cite{rahmaniani2017benders}  as the solution approach.
\end{itemize}

The remainder of this paper is organized as follows. Section \ref{sec:relatedwork} reviews the existing literature on rural communication systems, intent-based networking, and satellite-assisted IAB architectures. Section \ref{sec:systemmodel} presents the system model and details the proposed intent-aware satellite-assisted IAB-based FWA. Section \ref{sec:ProblemFormulation} discusses the problem formulation. Section \ref{sec:solution} describes the solution approach. Section \ref{sec:performancEevaluation} presents performance evaluations and discusses the effectiveness of the proposed framework. Finally, Section \ref{Conclusion} concludes the paper and outlines directions for future research.

\section{Literature Review}
\label{sec:relatedwork}
\emph{Rural Area Connectivity:}
Rural and remote regions often face persistent connectivity challenges due to low population density, difficult terrain, and limited infrastructure investment. Traditional cellular networks and fixed broadband solutions are economically unattractive for operators in these areas, resulting in digital exclusion \cite{ndikumana2025energy, yaacoub2020efficient}. To address this issue, FWA has been explored as a cost-effective alternative to fiber deployment. FWA enables broadband-like performance using wireless backhaul, but its coverage is often limited to a single hop between the base station and end users \cite{oproiu20185g, ndikumana2024renewable, ndikumana2024digital}. To extend coverage beyond FWA's capabilities, IAB has emerged as a promising 5G solution \cite{specification3gpp}. IAB enables multi-hop connectivity via wireless backhaul links, improving deployment flexibility and reducing reliance on fiber. However, in the IAB network, the donor still needs fiber-optic connectivity to access the internet. Consequently, maintaining reliable multi-hop links remains a technical challenge for rural deployments \cite{yaacoub2020efficient}.
\begin{figure*}[t]
	\centering	\includegraphics[width=2.0\columnwidth]{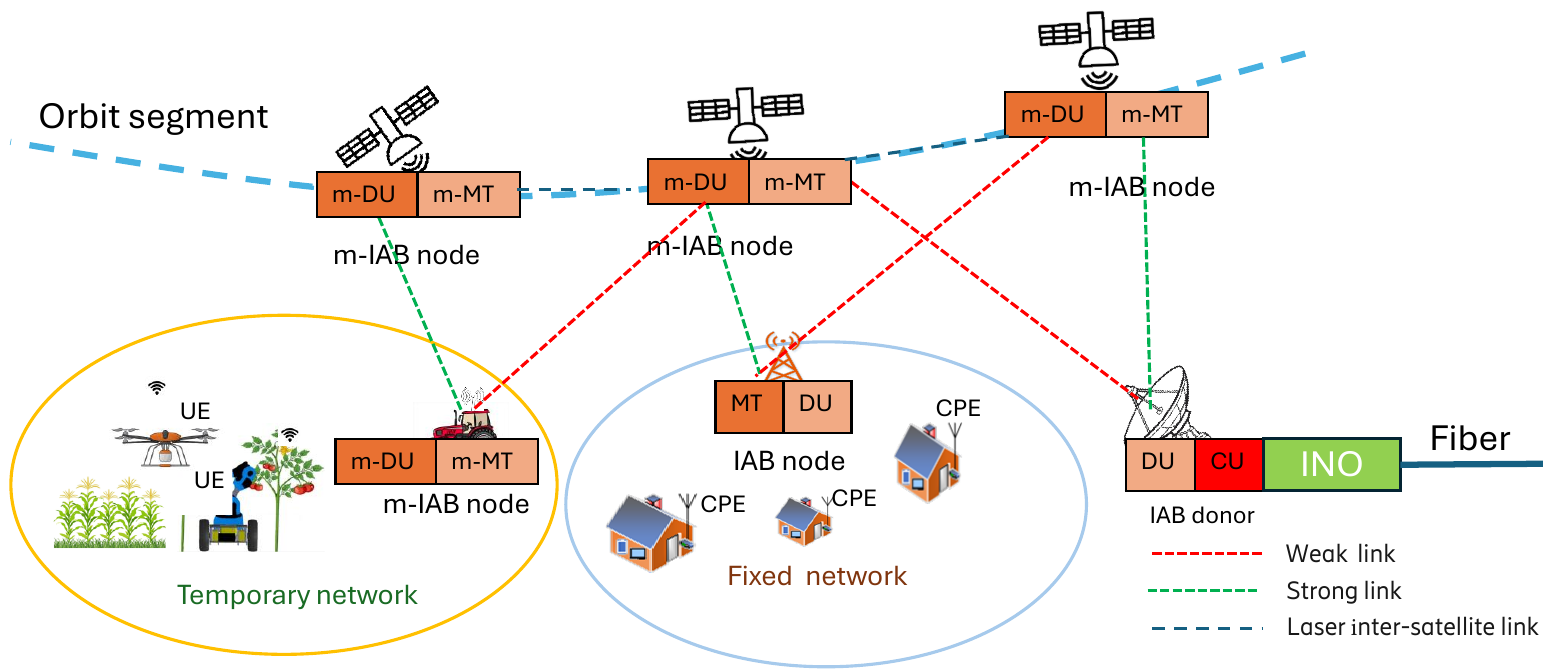}
	\caption{Illustration of our system model for a rural area.}
	\label{fig:systemmodel2}
\end{figure*}

\emph{Satellite-Assisted IAB Network:}
Satellite communications offer a complementary solution to the limitations of terrestrial backhaul. The integration of Low Earth Orbit (LEO) satellite constellations with terrestrial networks provides wide-area coverage, mobility, and global accessibility \cite{kodheli2020satellite}. Unlike traditional geostationary systems, LEO satellites offer lower latency and higher throughput, making them suitable for backhaul support in dynamic or temporary rural operations. Several studies have investigated satellite-assisted 5G networks \cite{majamaa2023satellite, xu2022quantum}, in which satellites serve as backhaul providers for remote base stations or IAB stations \cite{abdullah2023integrated, abdullah2024integrated}. However, while satellite-assisted IAB networks enhance coverage, they introduce new challenges, including link intermittency, increased handover complexity, and delay variations due to satellite motion.
Additionally, resource management across satellite and terrestrial segments requires joint optimization of radio, backhaul, and computational resources to achieve seamless service continuity. Furthermore, rural activities, such as seasonal farming \cite{kahmann2025nomadic} or mining using connected devices, create temporally varying connectivity demands that require adaptive network configurations.

\emph{Intent-based Networking for Dynamic Resource Allocation:}
The concept of IBN has gained attention as a paradigm that allows users or applications to express high-level goals or intents, which are automatically translated into low-level network configurations \cite{leivadeas2022survey}. Intent-based management frameworks have been applied in data centers and enterprise networks to simplify automation, improve agility, and reduce operational complexity. In mobile and edge networks, intents can express service requirements such as latency bounds, bandwidth guarantees, or reliability levels \cite{mehmood2023intent, wang2025survey}.
Recent studies have explored intent-aware orchestration for next-generation networks \cite{sidhu2025intent}. For example, \cite{yang2021automatic} proposed intent-driven network slicing to allocate resources based on service-level intents dynamically. Rural regions exhibit dynamic population movements and activity-dependent connectivity needs, which require context-aware intent interpretation and adaptive resource orchestration across terrestrial and non-terrestrial segments \cite{ramirez2025multi}. The coexistence of residential and field-based network demands in an intent-based network has not been explored in existing literature for rural areas.

\emph{Research Gap and Motivation:} Although prior studies have addressed FWA and IAB architectures, satellite-terrestrial integration, and intent-based management, the coordination between temporary field networks (e.g., during agricultural or forestry operations) and fixed residential networks to minimize energy consumption while satisfying data rate requirements remains largely unexplored. Moreover, current related works lack mechanisms that dynamically interpret user or operator intent to control when and where to activate network functions, scale capacity, and ensure energy efficiency. This gap motivates the development of an Intent-Aware Satellite-Assisted IAB-based FWA for rural areas. By leveraging user intents, such as farm operation schedules or machinery usage patterns, the network can proactively deploy temporary connectivity, dynamically scale backhaul links, and deactivate network functions when activities cease. This intent-driven coordination between satellite backhaul and IAB terrestrial nodes enables both energy-efficient and on-demand connectivity for rural operations. Overall, the proposed Intent-Aware Satellite-Assisted IAB-based FWA introduces a new paradigm for rural connectivity by combining intent-based automation, dynamic resource management, and hybrid satellite–terrestrial networking. It enables a flexible, context-driven, and energy-efficient communication network that supports both temporary field and residential broadband services.
\begin{figure}[t]
	\centering	\includegraphics[width=0.8\columnwidth]{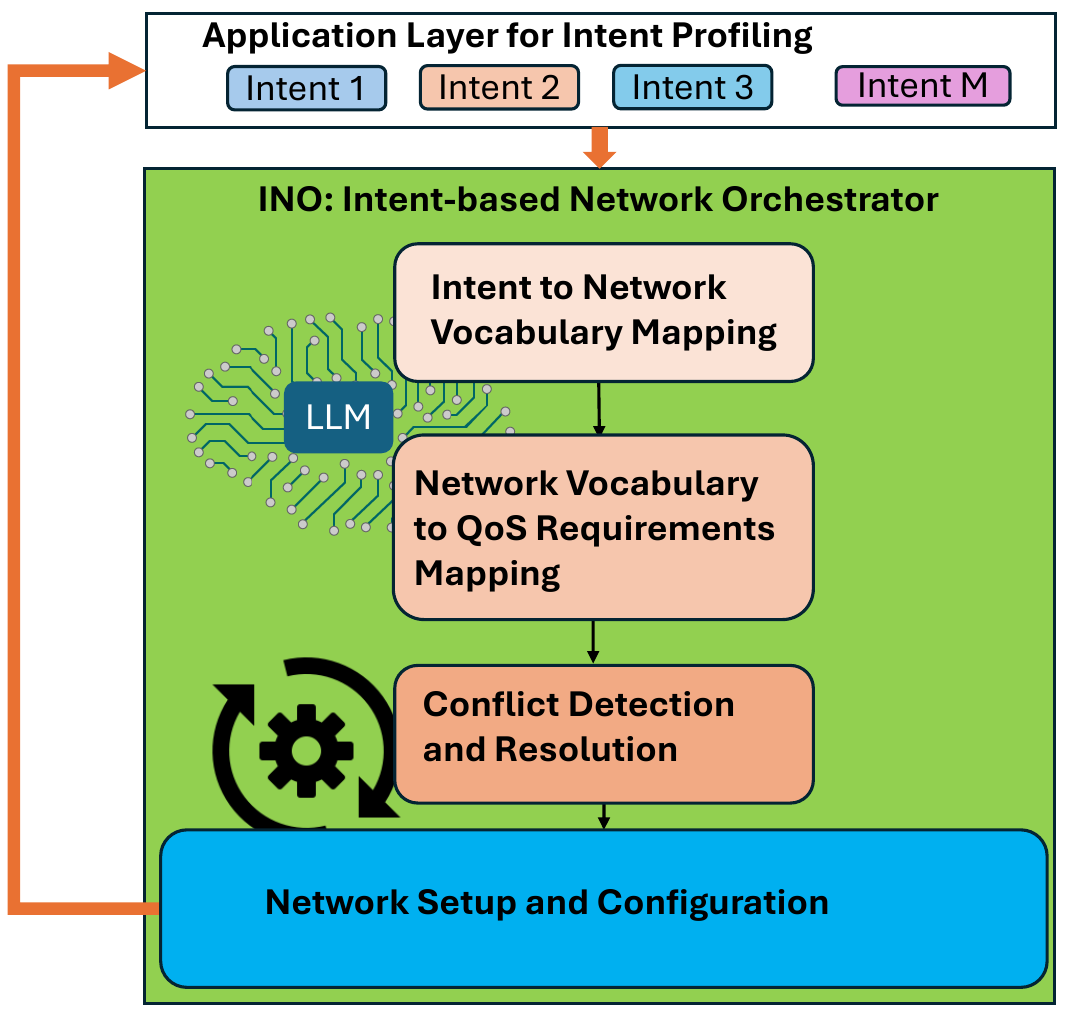}
	\caption{ Illustration of INO-Intent-based Network Orchestrator.}
	\label{fig:Intent}
\end{figure}
\section{System Model}
\label{sec:systemmodel}
\begin{table}[t]
	\caption{Summary of key notations.}
	\label{tab:table1}
	\begin{tabular}{ll}
		\toprule
		Notation & Definition\\
		\midrule
  $\mathcal{T}$& Time slots, $t=1,\dots,T$\\
 $\mathcal{U}$& Set of terminals ( i.e., UEs/CPEs) \\
     $\mathcal{S}$& Set of satellites IAB nodes \\
     $\mathcal{N}$& Set of terrestrial IAB nodes \\
    $\mathcal{E}$& Set of directed links \\
     $\mathcal{M}$& Set of user terminal intents\\
      $\mathcal{V}$ & A set of network vocabulary\\
    $\mathcal{Q}$ &  set of network QoS\\
	$m^u_t$& User intent at time $t$\\
     $d^u_t$ & Data rate requirement of terminal $u$\\ 
    $C^{i,j}_t$ & Link capacity between satellites
    $i$ and $j$ \\
     $C^{j,k}_t$ & Link capacity between terrestrial IAB station $k$  \\
   & with satellite m-IAB node $j$\\
    $s^e_t$& Satellite visibility  for forming link at time $t$\\
   $b_1$& Intent to network vocabulary mapping function\\
    $b_2$& Network vocabulary to QoS  mapping function\\
    $ \tilde{L}^{i,j} $& Delay for intersatellite link\\
    $x^u_t$& User admission variable \\
   $\tilde{L}_{\max}$& Maximum  delay\\
   $R_{\max}$& Maximum data rate\\
     $E^i$& Energy consumption at IAB node $i \in \mathcal{S} \cup \mathcal{N} $\\
      $a_t^i$&   MEC server activation decision variable at IAB node $i$\\
     $z^e_t$& Activation of link $e$ at time $t$\\
		\bottomrule
	\end{tabular}
\end{table}
In Fig. \ref{fig:systemmodel2}, we consider a system model with one gateway as an IAB donor, multiple   Low Earth Orbit (LEO) satellites as IAB nodes, and terrestrial fixed and mobile IAB nodes. In other words,  each LEO satellite has on-board processing, i.e., a regenerative satellite, that provides uplink and downlink services to terrestrial IAB nodes and the donor. We denote $\mathcal{S}$ as a set of satellite IAB nodes and $\mathcal{N}$ as a set of terrestrial fixed and mobile IAB nodes.   Each fixed IAB node comprises a DU and an MT unit. On the other hand, each mobile IAB node comprises a  mobile Distributed Unit (m-DU) and a mobile Mobile Termination (m-MT) unit. The DU/ m-DU serves both CPEs and downstream MTs/m-MTs. The MT/m-MT allows the node to operate as a relay when connecting to its parent DU/m-DU. Furthermore, the IAB donor includes both a DU and a CU. Unless stated otherwise, we use the term IAB station to refer to either an IAB node, m-IAB node,  or an IAB donor.
The terrestrial IAB stations serve user terminals, i.e., user devices (UEs) and Customer Premises Equipment (CPE), in rural areas under intent-driven demands. Unless stated otherwise, we use the term terminal to refer to either a UE or a CPE. We denote  $\mathcal{U}$ as a set of user terminals and $\mathcal{M}$ as a set of user intents. 

In serving user terminals $\mathcal{U}$ in rural areas, we consider the orbital segment. Depending on the location of the field area requiring temporary network access and the residential areas requiring an FWA network, either circular or elliptical orbits can be considered. For instance, in polar regions, elliptical orbits are preferred over circular ones because their high inclination enables them to pass over the poles, providing better coverage. In this work, we focus on a segment of the orbit \cite{ullah2025extending} and  LEO  satellites. Here, we remind you that the LEO satellite can have an elliptical orbit, in which the satellite altitude varies between perigee (the closest point) and apogee (the farthest point). If both the perigee and apogee altitudes of the satellite lie within the LEO range ($100$-$2000$ km), then the satellite is considered to be in a LEO elliptical orbit. The choice between a circular and an elliptical orbit for the IAB station in our solution can be made based on the rural area's geographic location.
Furthermore, we consider $\mathcal{T}$ to be the period during which the satellite travels in the rural area and is slotted. Therefore, we define $s^e_t\in\{0,1\}$ as visibility of satellite $s \in \mathcal{S}$ for forming link $e \in \mathcal{E}$ at time $t$ during the pass arc, where $\mathcal{E}$ is the set of directed links. The key notations used in this paper are shown in Table \ref{tab:table1}.           

\section{Problem Formulation}
\label{sec:ProblemFormulation}
As shown in Fig. \ref{fig:Intent}, in this section, we discuss the problem formulation for Intent-to-network QoS mapping and conflict resolution, and optimization to maximize energy efficiency for establishing the network shown in Fig. \ref{fig:systemmodel2}.

\subsection{Intent Profiling}
Here, we assume that the FWA exists in rural areas and intent helps to improve it and establish a temporary network. For establishing temporal and improving FWA network shown in Fig. \ref{fig:systemmodel2}, we assume each user of a terminal $u \in \mathcal{U}$  in residential expresses an intent at an abstract level for time $t$ and location $l$, denoted by $m^u_{t} \in \mathcal{M}$,
where $\mathcal{M}= \{1, 2, \dots, M\}$ represents the set of all possible intents. The intents are sent to the Intent-based Network Orchestrator (INO) via the application layer. This process is called intent profiling. Here are some examples of user intents:
\begin{itemize}
	\item User 1: Real-time monitoring of soil moisture sensors with minimal energy usage for a specific location and time.
	\item User 2: Streaming HD farm activities to the farm office, farm location, and time.
	\item User 3: Secure remote control of mining equipment, mining site location, and time.
\end{itemize}

\subsection{Intent to Network Vocabulary Mapping}
We assume the INO has a network vocabulary, where $\mathcal{V} = \{1, 2, \dots, V\}$ denotes the set of vocabulary elements. The following are examples of the network vocabulary: $v_1$: low latency,  $v_2$: energy efficiency, $v_3$: high throughput,  $v_4$: secure connection, $v_5$: reliable connectivity.

When intent reaches the INO at the IAB donor, the INO must search the network vocabulary that matches the user's intent to understand the user's requirements. Therefore, each user intent needs to be matched with  one or more network vocabulary elements using the following mapping function:
\begin{equation}
	b_1 : \mathcal{M} \rightarrow \mathcal{V}.
\end{equation}
For each user terminal $u \in \mathcal{U}$, the corresponding network vocabulary elements for its intent $m^u_{t}$ can be expressed as:
\begin{equation}
	V^u_{t} = b_1(m^u_{t}), 
\end{equation}
where $V^u_{t} \subseteq \mathcal{V}$ represents the subset of network vocabulary elements associated with intent $m^u_{t}$. This is an example of intent-to-network-vocabulary mapping using $b_1$:
\begin{itemize}
	\item User 1: $\{v_1, v_2, v_5\}$ (low latency, energy efficiency, reliable connectivity).
	\item User 2: $\{v_3, v_5\}$ (high throughput, reliable connectivity).
	\item User 3: $\{v_1, v_4, v_5\}$ (low latency, secure connection, reliable connectivity).
\end{itemize}

\subsection{Network Vocabulary to QoS Requirements Mapping}
We consider $\mathcal{Q} = \{1, 2, \dots, Q\}$ as the set of network Quality of Service (QoS) needed in rural areas, 
such as latency, throughput, reliability, and energy efficiency. 
Each network vocabulary element $v \in \mathcal{V}$ is mapped to one or more QoS parameters by the following function:
\begin{equation}
	b_2 : \mathcal{V} \rightarrow \mathcal{Q}.
\end{equation}
Hence, for an extracted vocabulary element $v$ from an intent, the associated QoS parameters are $
Q_{v} = b_2(v)$,
where $Q_{v} \subseteq \mathcal{Q}$. Here, we assume that the network provider has the network vocabulary-QoS requirements mapping.

For each user terminal $u \in \mathcal{U}$, the overall QoS requirements can be obtained by aggregating the QoS sets corresponding to all vocabulary elements derived from the user’s intent $m^u_{t}$ as follows: 
\begin{equation}
	Q^u_{t}= \bigcup_{v \in V^u_{t}} b_2(v).
\end{equation}
This is an example of intent-to-network QoS mapping using b2:
\begin{itemize}
	\item User 1: $\{q_1, q_2, q_5\}$ (Latency $\leq 50$ ms, Packet loss $\leq 1\%$, Uptime $\geq 99\%$ ).
	\item User 2: $\{q_3, q_2, q_5\}$ (Throughput $\geq 20$ Mbps, Packet loss $\leq 1\%$, Uptime $\geq 99\%$).
	\item User 3: $\{q_1, q_4, q_4, q_5\}$ (Latency $\leq 50$ ms, Encryption enabled, Packet loss $\leq 1\%$, Uptime $\geq 99\%$).
\end{itemize}

\subsection{Composite Mapping of Intent to QoS Requirements}
By combining the above mapping functions $b_1$ and $b_2$, we define a composite function that directly associates user intents with network QoS requirements:
\begin{equation}
	b = b_2 \circ b_1 : \mathcal{M} \rightarrow \mathcal{Q}.
\end{equation}
Therefore, for each user terminal $u \in \mathcal{U}$, the final QoS requirement derived from its intent $m^u_{t}$ is given by
\begin{equation}
	b(m^u_{t})  = \bigcup_{v \in b_1(m^u_{t})} b_2(v).
\end{equation}

\subsection{Conflict Detection and Resolution}

Conflicts may arise when multiple users simultaneously request network services with QoS requirements that cannot be jointly satisfied within the same time interval at the same location in a rural area. To model this, we define a conflict-free indicator function between any two users $u \in \mathcal{U}$ and $w \in \mathcal{U}$ as:
\begin{equation}
	c(u,w) =
	\begin{cases}
		1, \; \text{if both $b(m^u_{t}) $ and $b(m^w_{t}) $}\\ 
		\; \; \; \text{ can be simultaneously satisfied,} \\
		0, \;  \text{otherwise.}
	\end{cases}
\end{equation}

To handle the above conflicts, we introduce a binary user-admission variable $x^u_t$ for each user $u$, where $x^u_t=1$ indicates that the user's intent can be satisfied in the current scheduling period. Otherwise, $x^u_t=0$. Since the network's QoS requirements are numerous, we focus here on delay and data rate requirements. Other requirements, such as packet loss and error rate, will be considered for future work. Then, we formulate the following user admission optimization as a $0$-$1$ integer linear program:
\begin{equation}
\label{eq:conflict_corrected}
\begin{aligned}
\max_{\vect{x}} \quad 
& \sum_{u=1}^{U}x^u_t \\
\text{s.t.} \quad 
& \sum_{u=1}^{U} d^u_t x^u_t \le R_{\max}, \quad \forall t,\\
& \sum_{u=1}^{U}
L^{u}_t x^u_t \le \tilde{L}_{\max}, \quad \forall u,\\
& x^u_t + x^w_t\le 1 + c(u,w), \quad\forall\ 1\le u < w\le U,\\
& x^u_t, x^w_t \in \{0,1\}, \quad \forall u, w \in \mathcal{U}.
\end{aligned}
\end{equation}
where $R_{\max}$ denotes the maximum data rate the network can provide at a time $t$ and $d^u_t$ is the data rate required for each user terminal $u$ based on intent analysis. Furthermore, we define $L^{u}_t$ as the latency requirement for each user terminal $u$ based on intent analysis. $\tilde{L}_{\max}$ represents the maximum tolerable end-to-end latency, accounting for all user terminals that need to be connected to the network.

The constraint $x^u_t + x^w_t\le 1 +c(u,w)$ ensures that conflicting users are not admitted simultaneously. This means that if users $u$ and $w$ are in conflict, i.e., $c(u,w)=0$, they cannot both be active at the same time; one must be denied based on the time of arrival of the intent. In the absence of conflict, both users can be accepted. The objective function maximizes the total number of users that the network can admit, subject to network capacity and end-to-end latency constraints. 

\subsection{Network Setup and Configuration}
After identifying the total number of users the network can accommodate and the network's QoS requirements, the next step is to set up and configure the network. We consider satellites as mobile IAB nodes, and terrestrial mobile and fixed IAB nodes, where m-DU, m-MT, DU, and MT can be deployed as Virtual Network Functions (VNFs) \cite{vieira2026virtual}. VNFs require a computer to
host them. Here, VNFs are software implementations that run on general-purpose computing
hardware, often referred to as commercial off-the-shelf (COTS) servers. Here, we consider a small Multi-access Edge Computing (MEC) server. In other words, VNFs can be activated when needed. When they are not needed, they can be disabled to save energy. Each satellite with active VNFs can be assigned a specific area called a footprint, based on its orbit segment. Here, we assume the satellite's orbit segment is known and predictable. During the satellite pass, the satellite can serve terrestrial mobile and fixed IAB nodes within its footprint.

We model the power consumption at IAB station $i$ at time $t$ as:
\begin{equation}
	\varphi^i_t = \Upsilon_{\mathrm{base}}^i +  \Upsilon^i_t  \nu^{i}_t,
\end{equation}
where $\Upsilon_{\mathrm{base}}^i$ is the baseline power of IAB station $i\in\mathcal{S}\cup\mathcal{N}$. 
We define $\Upsilon^i_t$ as the power of activating the MEC server with VNFs at time $t$ such that:
\begin{equation}
	\Upsilon^i_t=\psi_{base}+\sum_{\chi=1}^{\tilde{\chi}}  \psi^{i}_\chi,
\end{equation}
where $\tilde{\chi}$ denotes the number of VNFs deployed at the MEC server, including m-DU, m-MT, DU, and MT. We define $\psi^{i}_{\chi}$ as the power consumption associated with an active VNF, while $\psi_{\text{base}}$ represents the baseline power consumption when the MEC server is powered on but not processing any traffic. Furthermore, the power consumption of VNF depends on the number of terminals connected to it. Therefore, we define $\nu^i_t$ as the traffic load at VNFs deployed in the MEC server at IAB station $i$  such that:
\begin{equation}
	\nu^{i}_t = \sum_{u=1}^{U^i} \frac{ \tilde{\lambda}_t^{i, v}}{ \tilde{\mu}_t^{i, v}},
\end{equation}
where $U^i$ is the number of user terminals connected to IAB node $i$, $ \tilde{\lambda}_t^{i, v}$ is packet arrival rate, and $ \tilde{\mu}_t^{i, v}$ is service rate at time $t$.

We define a decision variable  $a_t^i\in\{0,1\}$ for activating an  MEC server with VNFs at IAB node $i\in\mathcal{S}\cup\mathcal{N}$ in slot $t$, where $\Delta t$ is the slot duration, such that:
\begin{equation}
	a_t^i =
	\begin{cases}
		1, \; \text{if  $\nu^{i}_t \neq 0$, activate MEC server with
			VNFs,} \\
		0, \;  \text{otherwise.}
	\end{cases}
\end{equation}
Then, we define the energy consumption of IAB node $i$ as follows:
\begin{equation}
	E^i=( \Upsilon_{\mathrm{base}}^i + a_t^i \Upsilon^i_t  \nu^{i}_t)  \Delta t.
\end{equation}
In other words, when IAB functions are not activated, there is no need to use the MEC server. Therefore, when $a_t^i=0$,  $ a_t^i \Upsilon^i_t  \nu^{i}_t  \Delta t=0$ and the energy consumption when the IAB node with MEC server and VNFs disabled in timeslot $t$ becomes 
$E^i= \Upsilon_{\mathrm{base}}^i\Delta t$.

\emph{Inter-satellite backhaul}:
Let us consider two interconnected LEO satellites, i.e., m-IAB nodes, denoted as 
$i$ and  $j$. We assume these satellites maintain perfect antenna alignment and that the intersatellite link experiences negligible interference from signals transmitted by other preceding or succeeding LEO satellites. Consequently, the intersatellite link capacity between  the LEO satellites
$i$ and $j$ can be expressed as follows:
\begin{equation}
	C^{i,j}_t  =  B^{i,j}\log_2\!\Big(1+\frac{P^i_t (D^{i,j})^{-\varkappa} G^j_t}{\sigma^2}\Big),\forall e=(i \to j) \in \mathcal{E}
\end{equation}
where  $P^i_t$ is the transmission power when the IAB node with MEC server and VNFs is active. $D^{i,j}$ is the distance between the satellites $i$ and $j$ and $-\varkappa$ represents the path loss exponent.  Furthermore,
we define de delay $ \tilde{L}^{i,j} $ for intersatellite link as:
\begin{equation}
	 \tilde{L}^{i,j} =  \frac{\sum_{u\in\mathcal{U}^{i,j}} x^u_t d^u_t \Delta t}{C^{i,j}_t }
\end{equation}
where $\mathcal{U}^{i,j}$ is the set of terminals using intersallelite link between satellites  $i$ and $j$ for transmission duration $\Delta t$.
\begin{figure}[t]
	\centering	\includegraphics[width=1.0\columnwidth]{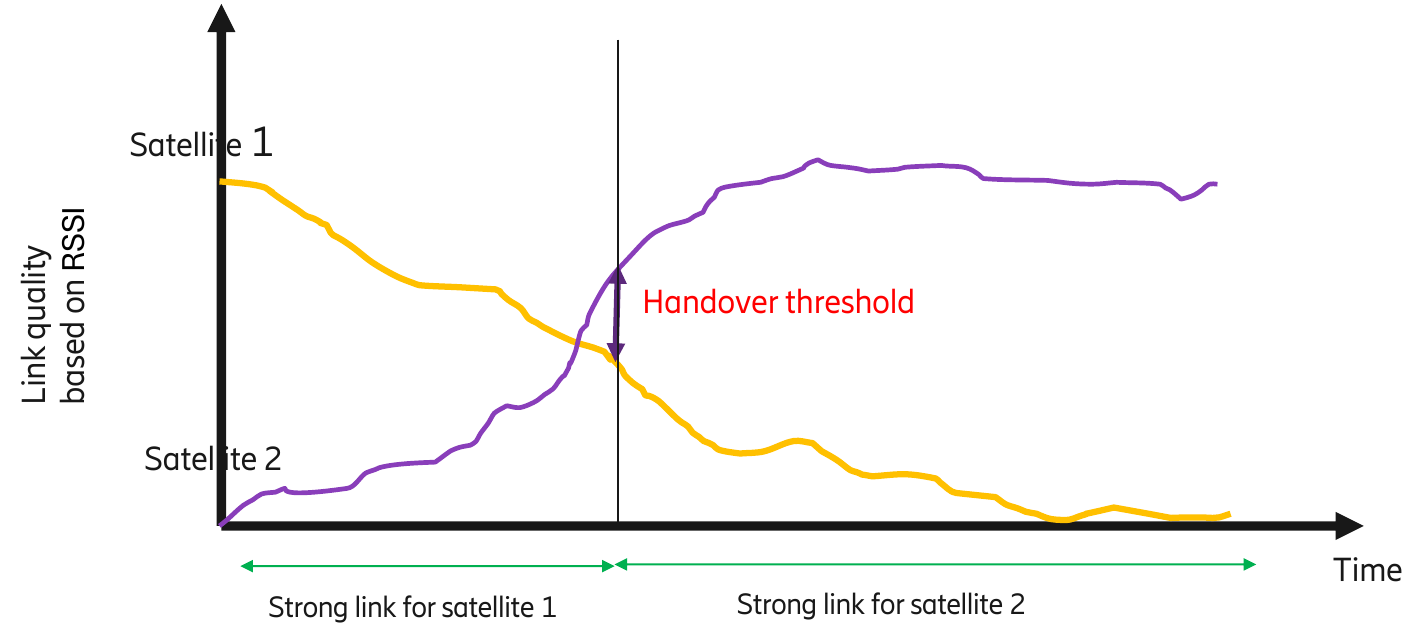}
	\caption{Link quality and handover illustration.}
	\label{fig:Linkhandover}
\end{figure}

\emph{Link between satellite IAB and terrestrial IAB nodes: }
We assume that satellite passes are known and predictable based on publicly available satellite orbital datasets, such as those provided by CelesTrak \cite{satellite}. As shown in Fig. \ref{fig:Linkhandover}, the terrestrial IAB station, such as the IAB node, m-IAB node, and IAB donor, continuously measures the received signal strength indicator (RSSI) of available satellites in its area. At any given time, the link with the highest RSSI is selected as the strong link, while the remaining links are maintained as weak links. When the RSSI of the current strong link falls below a predefined handover threshold, the IAB station reclassifies this link as a weak link and promotes the link with the highest RSSI to become the new strong link for handover. This dynamic link selection mechanism is similar to the handover mechanism discussed in \cite{yang2016seamless}.
Furthermore, IAB stations need to inform the IAB donor about its available links so that the IAB donor maintains details of all strong and weak links to reach each IAB station. This mechanism enables seamless connectivity and improves link robustness under time-varying satellite visibility conditions. Therefore, we consider the handover delay $\tilde{L}_h^{j,k}$ between  terrestrial IAB station $k$  with satellite $j$ (m-IAB node), which is defined as the delay associated with the handover process and is given by:
\begin{equation}
	\tilde{L}_h^{j,k} = (1 - P_h) T_r + P_h T_h ,
\end{equation}
where \( P_h \) denotes the probability of a successful handover. Specifically, \( P_h = 1 \) if the handover succeeds and \( P_h = 0 \) otherwise. The parameter \( T_h \) represents the time required to complete a successful handover, while \( T_r \) denotes the additional time incurred to retry the handover in the event of failure. Furthermore, we define the capacity of the link between terrestrial IAB station $k$ and satellite m-IAB node $j$ as follows:
\begin{equation}
	C^{j,k}_t = B^{k,j} \log_2 \left( 1 + \frac{P_t^j \left(D^{j,k}\right)^{-\varkappa} G_t^{k}}{\sigma^2} \right),
\end{equation}
where $D^{j,k}$ is the distance between terresterial IAB station $k$ with satellite $j$.  The distance $D^{j,k}$ can be expressed as follows:
\begin{equation}
	D^{j,k}=  \frac{(\kappa^j + \xi^j) \sin \beta^j}{\cos \theta^j}
\end{equation}
where $\kappa^j$ is Earth's radius and $\xi^j$ is the orbit height, i.e., the altitude from a point in a rural area directly below satellite $j$. We define $\beta^j$ as the coverage angle and $\theta^j$ as the elevation angle.

Considering  the capacity of the link $C^{j,k}_t$ and  terminals using terrestrial IAB station $k$ connected to satellite $j$, we can define delay as follows:
\begin{equation}
	\tilde{L}_h^{j,k}=  \frac{\sum_{u\in\mathcal{U}^{j,k}} x^u_t d^u_t\Delta t}{C^{j,k}_t},
\end{equation}
where $\mathcal{U}^{j,k}$ is the  user terminals connected to satellite $j$ via IAB station $k$ in transmission duration $\Delta t$.

\emph{Access link between terrestrial IAB station and terminals:}
The capacity of the  access link between terminal $u$ and terrestrial IAB station $k$ is defined as follows:
\begin{equation}
	C^{u,k}_t  = a_t^k B^u\log_2\!\Big(1+\frac{P^k_t G^u_t}{\sigma^2}\Big),
\end{equation}
where $B_u$ is the channel bandwidth, $G^u_t$ is the channel gain, and  $\sigma^2$ is the noise power. Furthermore, we define the latency of the access link as follows:
\begin{equation}
\tilde{L}^{u,k}=  \frac{ x^u_t d^u_t\Delta t}{C^{u,k}_t}.
\end{equation}

We formulate the following optimization problem to maximize energy efficiency, specifically the data rate per energy consumption:
\begin{subequations}
	\label{eq:problem_formulation1}
	\begin{align}
		&
		\max_{a_t^i, s^e_t, d^u_t} \quad 
		\frac{\sum_{u \in \mathcal{U}} d^u_t}
		{\sum_{i\in\mathcal{S}\cup\mathcal{N}} E^i
		} \tag{\ref{eq:problem_formulation1}}\\
		\text{s.t.} \quad 
		& C^{u,k}_t  \ge x^u_t d^u_t, \forall u \in \mathcal{U}, t \in \mathcal{T}, \label{first:a1}\\
		& \sum_{u\in\mathcal{U}^{j,k}} x^u_t d^u_t \leq s^e_t C^{j,k}_t\quad \forall \; t\in\mathcal{T}, \label{first:a2}\\
		& \sum_{u\in\mathcal{U}^{i,j}} x^u_t d^u_t \leq s^e_tC^{i,j}_t,\forall e \in \mathcal{E}, t \in \mathcal{T}, \label{first:a3}\\
		& s^e_t \le a_t^i, \; s^e_t \le a_t^j, \forall e=(i \to j) \in \mathcal{E}, \label{first:a4}\\
		& \sum_{u=1}^{U}
		\tilde{L}^{u,k}+  \tilde{L}_h^{j,k}+ \tilde{L}^{j,k} +  \tilde{L}^{i,j}  \le \tilde{L}_{\max},\forall u \in \mathcal{U} \label{first:a5}.
	\end{align}
\end{subequations}
The constraints in (\ref{first:a1}), (\ref{first:a2}), and (\ref{first:a3}) ensure that the links meet the user terminals' data rate requirements, given the satellites' visibility. Constraint (\ref{first:a4}) guarantees that the satellites are activated as m-IAB nodes when they are visible. Constraint (\ref{first:a5}) ensures that the total latency is satisfied.

\section{Solution Approach}
\label{sec:solution}
In this section, we present a detailed solution approach for intent-to-QoS mapping, conflict resolution, and the optimization problem to maximize
energy efficiency.  
\subsection{LLM and Optimization Approach  for Intent-to-QoS Mapping and Conflict Resolution}

The mapping functions $
b_1 : \mathcal{M} \rightarrow \mathcal{V}$ and $
b_2 : \mathcal{V} \rightarrow \mathcal{Q}$  are unknown because the intents are expressed at an abstract level using natural language. Therefore, we can use AI models, such as LLMs \cite{chang2024survey}, to learn these functions from data. We choose LLMs over other AI frameworks because they can analyze user intents expressed at an abstract level by interpreting natural-language intents and translating them into executable network configurations or policies \cite{tu2025intent}. In other words, the LLM model can help us automatically learn how user intents map to network vocabulary and corresponding QoS requirements.

For $b_1: \mathcal{M} \rightarrow \mathcal{V}$, the first LLM learns a semantic mapping between user intents and the network vocabulary. Given an intent expressed as a high-level operational objective, the process begins with sentence segmentation. Specifically, each sentence $g^u$ from the user intent $m_t^u$ and each vocabulary term $v \in \mathcal{V}$ are encoded into embedding vectors using a pre-trained sentence embedding model $h_1(\cdot)$, yielding sentence embedding $\mathbf{o}^u = h_1(g^u)$ and network vocabulary embedding $\mathbf{w}^u = h_1(v)$. To mitigate LLM hallucinations during the semantic mapping process, we employ a pre-trained sentence-level LLM and then fine-tune the model. In the fine-tuning process, we consider positive and negative training pairs $(g^u, v, y_{gv})$, where $y_{gv}=1$ if the network vocabulary term $v$ is semantically associated with sentence $g^u$ (positive pair), and $y_{gv}=0$ otherwise (negative pair). The pre-trained LLM is fine-tuned by minimizing the following cosine-similarity-based loss function:
\begin{equation}
	L_1 = \frac{1}{|\mathcal{\tilde{V}}|} \sum_{(g^u, v) \in \mathcal{\tilde{V}}} \left(\text{cos\_sim}(\mathbf{o}^u, \mathbf{w}^u) - y_{gv}\right)^2,
\end{equation}
where the cosine similarity between embeddings is defined as:
\begin{equation}
	\text{cos\_sim}(\mathbf{o}^u, \mathbf{w}^u) = \frac{\mathbf{o}^u \cdot \mathbf{w}^u}{|\mathbf{o}^u| |\mathbf{w}^u|}.
\end{equation}
$\mathcal{\tilde{V}}$ is the set of all sentences and network vocabulary pairs.

Once the model is fine-tuned, each sentence $g^u$ is mapped to the most relevant vocabulary terms $v$ by computing $\text{cos\_sim}(\mathbf{o}^u, \mathbf{w}^u)$ and minimizing the cosine similarity loss function.  This approach ensures that both semantically meaningful and quantitatively precise mappings are obtained between user intents and network vocabulary.

For $b_2 : \mathcal{V} \rightarrow \mathcal{Q}$,  the second LLM learns a semantic mapping between network vocabulary and QoS vocabulary. In this work, we consider a semantic mapping from extracted network-related vocabulary in intent to a predefined QoS vocabulary. We use a pre-trained embedding LLM model $h_2(\cdot)$ with extracted network vocabulary embedding $\mathbf{w}^u = h_2(v)$ and  QoS requirement embedding $\mathbf{q}_u = h_2(q)$. Then, for LLM fine-tuning, we minimize cosine similarity loss defined as follows:
\begin{equation}
	L_2 = \frac{1}{|\mathcal{\tilde{Q}}|} \sum_{(v,q) \in \mathcal{\tilde{Q}}} \big( \text{cos\_sim}(\mathbf{w}^u, \mathbf{q}_u) - y_{vq} \big)^2
\end{equation}
where $\mathcal{\tilde{Q}}$ is the set of all network vocabulary and QoS vocabulary pairs. The $y_{vq}=1$ if the extracted network vocabulary term $v$ is semantically associated with QoS term $q$ (positive pair), and $y_{vq}=0$ otherwise (negative pair). For example, if the extracted network vocabulary lower-delay is matched with the lower-latency QoS vocabulary, $y_{vq} = 1$. Suppose lower delay is matched with throughput, $y_{vq} = 0$. In other words,  $y_{vq} = 1$ if similar meaning, i.e.,  high similarity. For a different meaning, i.e., lower similarity, $y_{vq} = 0$. Similarity between embeddings $v$ and $q$ is measured using the cosine similarity $\text{cos\_sim}(\mathbf{w}^u, \mathbf{q}_u)$. For example if $v$ is lower delay and $q$ is latency,  cosine similarity is $\text{cos\_sim}(\mathbf{w}^u, \mathbf{q}_u)=0.85$ and $l_{vq} = 1$. The loss function is $L_2=(0.85-1)^2=0.0225$. This is a good mapping because the $0.0225$ loss is too small.
After fine-tuning the LLM, the embeddings become domain-specific for the network, improving semantic alignment for QoS terms. Finally, we formulate end-to-end intent-to-QoS mapping, where we can learn the composite mapping function $b = b_2 \circ b_1: \mathcal{M} \rightarrow \mathcal{K}$
by minimizing the following loss function:
\begin{equation}
	\label{eq:llmloss}
	L = \lambda_1 L_1 + \lambda_2 L_2.
\end{equation}
Here, $\lambda_1>0$ and $\lambda_2>0$ are the weights to balance the contributions of each loss function.

After end-to-end intent-to-QoS mapping, we minimize (\ref{eq:conflict_corrected}), a binary ($0$-$1$) linear program that can be solved with modern mixed-integer programming (MIP) solvers such as Gurobi or CPLEX. Although NP-hard in general, such solvers employ branch-and-bound and cutting-plane techniques that can efficiently handle such formulated optimization problems. Furthermore, the LLMs and optimization approach for intent-to-QoS mapping and conflict resolution needed to be implemented in the INO available at the IAB donor. In other words, the IAB donor is always active.

\subsection{Solution for Energy Efficiency Maximization Problem}

In network setup and configuration, we need to solve the formulated optimization problem  (\ref{eq:problem_formulation1}) to maximize energy efficiency. The formulated problem (\ref{eq:problem_formulation1}) combined node and link activation with traffic routing, where traffic routing uses activated nodes and links. Therefore, node and link activation should start, followed by traffic routing. This motivates us to simplify and decompose the formulated problem. To solve  (\ref{eq:problem_formulation1}), we first simplify the objective function and then propose a practical two-stage Bender decomposition algorithm as a solution. In other words, we first reformulate the objective and then develop a two-stage Benders decomposition framework that separates node and link activation decisions from traffic routing.

To simplify the objective function, let us assume that, for each time slot $t$, the per-terminal data rate demand $d_t^u$ can be satisfied whenever terminal $u$ is admitted, and that the admission variable $x_t^u$ is obtained from solving (\ref{eq:conflict_corrected}). In other words, if the terminal is not admitted, it will not be connected to the network. The achievable data rate of terminal $u$ is therefore
\begin{equation}
	C_t^{u,k} = x_t^u d_t^u, 
	\qquad \forall u \in \mathcal{U}, \; \forall t .
\end{equation}
The total system throughput at time slot $t$ becomes
\begin{equation}
	R_t^{\mathrm{tot}} 
	\triangleq \sum_{u \in \mathcal{U}} C_t^{u,k}
	= \sum_{u \in \mathcal{U}} x_t^u d_t^u .
\end{equation}
Under this assumption, maximizing the ratio, i.e., energy efficiency with a known numerator, is equivalent
to minimize the denominator, i.e., total energy consumption for each time slot:
\begin{subequations}
	\label{eq:problem_formulation4}
	\begin{align}
		&\min_{\{a_t^i\}} 
		\sum_{i\in\mathcal{S}\cup\mathcal{N}} E^i
		\tag{\ref{eq:problem_formulation4}}\\
		& \text{subject to: (\ref{first:a1}) - (\ref{first:a5})}.
	\end{align}
\end{subequations}

To minimize energy consumption, we consider activating only a subset of nodes and links that can route all traffic demands while satisfying link capacity, satellite visibility, and latency constraints, rather than keeping all nodes and links active at all times. To obtain a self-contained formulation, we introduce a multi-commodity flow variable $f_t^{e,u} $. For each terminal $u \in \mathcal{U}$, directed link $e=(i \rightarrow j)$, and time slot $t$, let
$f_t^{e,u} \ge 0$ denote the flow rate of terminal $u$ over link $e$. The aggregate data frafic on link $e$ is $\sum_{u \in \mathcal{U}} f_t^{e,u}$. For a fixed time slot $t$, the problem can be formulated as the following Mixed-Integer Linear Programming (MILP):
\begin{subequations}
\label{eq:milp_full}
\begin{align}
 &\underset{(\vect{a}, \vect{s}, \vect{z},\vect{f} )}{\min} 
	\sum_{i\in\mathcal{S}\cup\mathcal{N}} E^i
	\tag{\ref{eq:milp_full}}\\
\text{s.t} \quad \nonumber\\
& \sum_{u \in \mathcal{U}} f_t^{e,u}
\le  C_t^{u,k} s_t^{e} z^e_t,\quad \forall e,
\label{eq:milp_capacity} \\
& \sum_{e:(i \rightarrow j)} f_t^{e,u}
- \sum_{e:(j \rightarrow i)} f_t^{e,u}
=
\begin{cases}
d_t^u, \quad i = \mathrm{src}(u), \\
- d_t^u, \quad i = \mathrm{dst}(u), \\
0, \quad \text{otherwise},
\end{cases}
&& 
\label{eq:milp_flow} \\
& \sum_{e:(j \rightarrow k)} f_t^{e,u}
- \sum_{e:(j\rightarrow k)} f_t^{e,u}
=
\begin{cases}
	d_t^u, \quad j = \mathrm{src}(u), \\
	- d_t^u, \quad j = \mathrm{dst}(u), \\
	0, \quad \text{otherwise},
\end{cases}
&& 
\label{eq:milp_flow2} \\
& \tilde{L}^{u,k}+  \tilde{L}_h^{j,k}+ \tilde{L}^{j,k} +  \tilde{L}^{i,j}  \le \tilde{L}_{\max},
\label{eq:milp_latency} \\
& s^e_t \le a_t^i, \; s^e_t \le a_t^j,  \label{eq:milp_visibility}\\
& z_t^{e} \le a_t^{i}, \quad
  z_t^{e} \le a_t^{j},\forall e=(i \to j) \in \mathcal{E},\label{eq:milp_activation} \\
& a_t^{i} \in \{0,1\}, \;
  z_t^{e} \in \{0,1\},\; s_t^{e}\in \{0,1\},\;
  f_t^{e,u} \ge 0,
\label{eq:milp_domain}
\end{align}
\end{subequations}
where $z^e_t$ is link $e$ activation decision variable at time $t$.

Problem \eqref{eq:milp_full} ensures the link is activated when the nodes that make the link are visible and activated. Then, traffic routing on the link can occur when the link is activated. In other words, \eqref{eq:milp_full}  jointly optimizes node/link activation and traffic routing.  We propose a two-stage Benders decomposition approach \cite{rahmaniani2017benders} that separates node and link activation decisions (the master problem) from traffic routing (the subproblem). We choose the two-stage Benders decomposition approach over other optimization methods due to its decomposition capability, scalability, and ability to handle uncertainty \cite{van2024converging}.

\emph{Stage I: Master Problem for Node and Link Activation}. For each terminal $u$, data-rate and latency requirements are known from intent analysis, and satellite visibility is predictable. Based on the terminal's location and satellite visibility, we determine candidate nodes and links to connect the terminals to the internet. We formulate the master problem  as follows:
\begin{subequations}
	\label{eq:master_problem}
	\begin{align}
		&\underset{(\vect{a}, \vect{s}, \vect{z} )}{\min} 
		\sum_{i\in\mathcal{S}\cup\mathcal{N}} E^i
		\tag{\ref{eq:master_problem}}\\
		\text{s.t.} \quad \nonumber\\
		& \sum_{u \in \mathcal{U}}
		d_t^u 
		\le C_t^{u,k} s_t^{e} z^e_t,
		\label{eq:master_capacity}\\
			& \sum_{u\in\mathcal{U}^{j,k}} x^u_t d^u_t \leq s^e_t C^{j,k}_t, \label{first:a22}\\
		& \sum_{u\in\mathcal{U}^{i,j}} x^u_t d^u_t \leq z^e_t s^e_tC^{i,j}_t, \label{first:a33}\\
		& s^e_t \le a_t^i, \; z^e_t s^e_t \le a_t^j,  \label{eq:milp_visibility2}\\
		& z_t^{e} \le a_t^{i}, \quad
		z_t^{e} \le a_t^{j},\forall e=(i \to j) \in \mathcal{E},\label{eq:milp_activation2} \\
		& a_t^{i} \in \{0,1\}, \;
		z_t^{e} \in \{0,1\},\; s_t^{e}\in \{0,1\}.
		\label{eq:milp_domain2}
	\end{align}
\end{subequations}                
The objective function in \eqref{eq:master_problem} minimizes node activation energy consumption, while the constraints ensure that node activations depend on satellite visibility and data rate requirements.

%---------------------------
\paragraph*{Stage II: Subproblem  for Routing Feasibility}
Given satellite visibility, activated nodes, and links $( \mathbf{a}=\{a_t^i\},
\mathbf{s}=\{s_t^e\},
\mathbf{z}=\{z_t^e\})$ obtained from Stage~I, in the Stage II,  (\ref{eq:milp_full}) becomes linear program (LP)   by only considering constraints in \eqref{eq:milp_capacity}, \eqref{eq:milp_flow}, \eqref{eq:milp_flow2},  and (\ref{eq:milp_latency}) to determine flows  $\mathbf{f}=\{f_t^{e,u}\}$ that respect link capacity and satisfy latency requirements. If the LP is infeasible, Benders feasibility cuts are generated and added to the master problem~\eqref{eq:master_problem}. The master problem is then re-solved. If the LP is feasible, the node and link activation and routing decisions obtained constitute an optimal solution to the original MILP. The overall two-stage Benders decomposition algorithm is summarized as follows:
\begin{algorithm}[H]
	\caption{Two-Stage Benders Decomposition for Energy-Efficient IAB Activation.}
	\label{alg:benders_updated}
	\begin{algorithmic}[1]
		\State \textbf{Preconditions:} Satellite visibility is predictable, LLMs and conflict resolution are performed to get admitted users' requirements;
		\State \textbf{Input:}  Demands $\{d_t^u\}$, capacities $\{C^{u,k}_t,C^{i,j}_t,C^{j,k}_t \}$, visibility $\{s_t^e\}$,  latency bound $\tilde{L}_{\max}$;
		\State Precompute candidate nodes and links by solving master MILP \eqref{eq:master_problem};
		\Repeat
		\State Solve routing (\ref{eq:milp_full}) LP  by only considering constraints in \eqref{eq:milp_capacity}, \eqref{eq:milp_flow}, \eqref{eq:milp_flow2},  and (\ref{eq:milp_latency});
		\If{LP infeasible}
		\State Generate Benders feasibility cuts and add to master;
		\EndIf
		\Until{LP feasible};
		\State \textbf{Output:} Variables $ \mathbf{a},
		\mathbf{s},
		\mathbf{z}, \mathbf{f}$  and optimal energy efficiency.
	\end{algorithmic}
\end{algorithm}

\begin{remark}Computational Complexity of the Proposed Approach is $O(n^3)$.
	\label{RA-TPM1} 
\end{remark}                               

The LLM-based mapping model $h_1(\cdot)$ and $h_2(\cdot)$ are pre-trained using datasets ${(M,V)}$ and ${(V,Q)}$, respectively. For LLM model with $\theta_1$ and $\theta_2$ parameters, the mapping has complexity $\mathcal{O}(\theta_1|\mathcal{V}| + \theta_2|\mathcal{Q}|)$, while the LLM model fine tuning  considering $\zeta$ epochs has complexity $\mathcal{O}(\zeta(\theta_1|\mathcal{V}| +  \theta_2|\mathcal{Q}|))$. In other words, during LLM inference, complexity reduces to $\mathcal{O}(\theta_1|\mathcal{V}| + \theta_2|\mathcal{Q}|)$, which scales linearly with model size. After LLM-based mapping, the conflict resolution problem is formulated as a binary linear program with $x^u_t$ as a binary variable. In the worst case, when the vector $\vect{x}=\{x^u_t\}$ becomes large, solving it has a computational complexity of $\mathcal{O}(2^n)$, where $n$ is the dimension of $\vect{x}$.

The proposed two-stage Benders decomposition splits (\ref{eq:milp_full}) into sub-problems for node/link action activation and routing sub-problems. Let $n$ denote the number of Benders iterations. The master problem, defined over  binary variable variables $( \mathbf{a}=\{a_t^i\},
\mathbf{s}=\{s_t^e\},
\mathbf{z}=\{z_t^e\})$, has worst-case complexity $\mathcal{O}(n^{3})$, when vectors  $\vect{a}$,  $\vect{s}$,and  $\vect{z}$ become large. The subproblem in Stage II, which is for routing feasibility, Given the satellite visibility, activated nodes and, links $(\vect{a},  \vect{s},  \vect{z})$ obtained from Stage~I, in the Stage II,  (\ref{eq:milp_full}) becomes linear program (LP)   by only considering constraints in \eqref{eq:milp_capacity}, \eqref{eq:milp_flow}, \eqref{eq:milp_flow2},  and (\ref{eq:milp_latency}) to determine flows $\vect{f}$. This has computation complexity  $\mathcal{O}((n)$. Overall, the total complexity of two-stage Benders decomposition is $\mathcal{O}\big(n^3 +n\big)$. The proposed approach significantly improves scalability by decoupling routing from node and link activation decisions.

By accounting for LLM inference, conflict resolution, and the two-stage Benders decomposition, the overall computational complexity of the proposed approach is given by
$
	\mathcal{O}(\theta_1|\mathcal{V}| + \theta_2|\mathcal{Q}|) + \mathcal{O}(2^n) + \mathcal{O}\big(n^3 + n\big).
$
Since the cubic term dominates for large \(n\), the overall computational complexity of the proposed method is \(\mathcal{O}(n^3)\). This level of complexity remains acceptable for rural deployment scenarios, where the number of terminals is relatively low compared to urban and sub-urban areas.

\section{Simulation Results and Analysis}
\label{sec:performancEevaluation}

\begin{figure}[t]
	\begin{minipage}{0.45\textwidth}
		\centering
		\includegraphics[width=1.0\columnwidth]{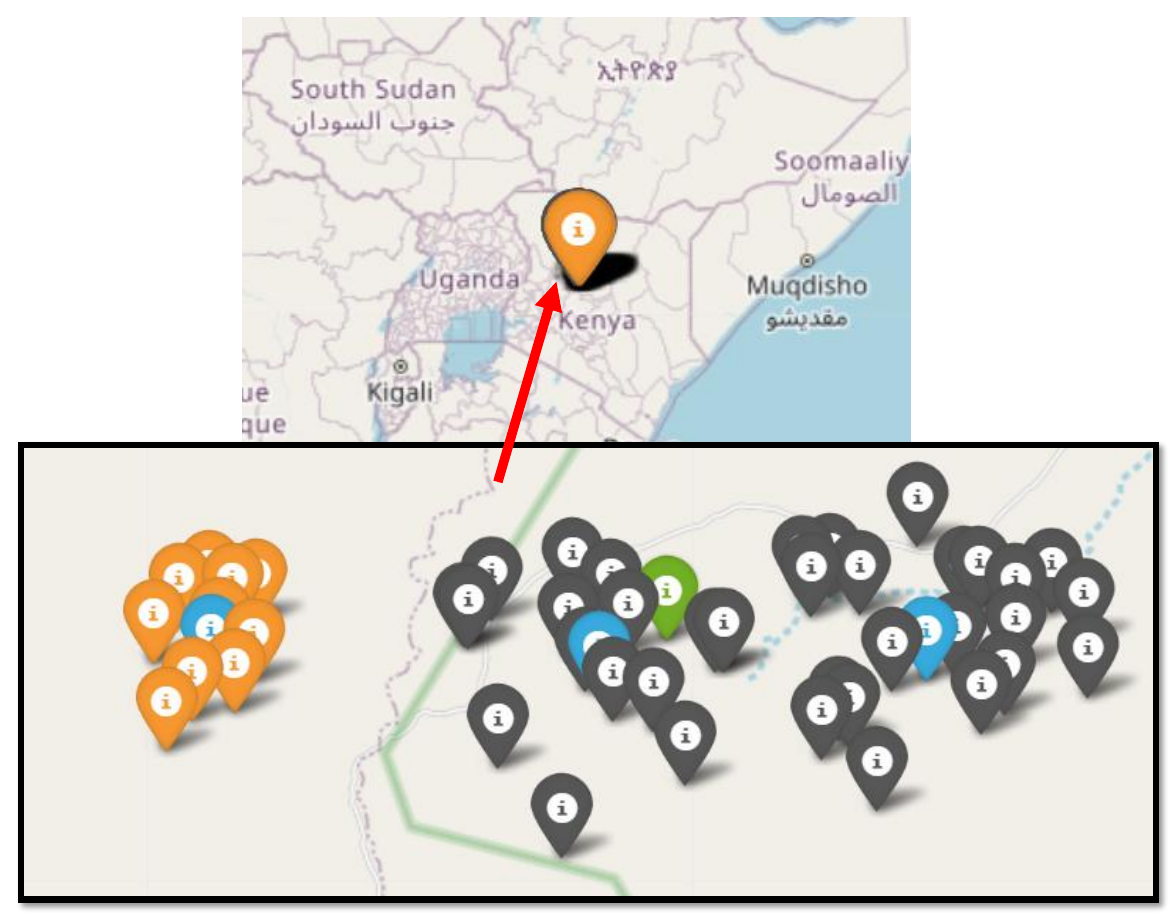}
		\caption{Selected region for rural areas.}
		\label{fig:Region}
	\end{minipage}
	\centering
	\begin{minipage}{0.45\textwidth}
		\centering
		\includegraphics[width=1.0\columnwidth]{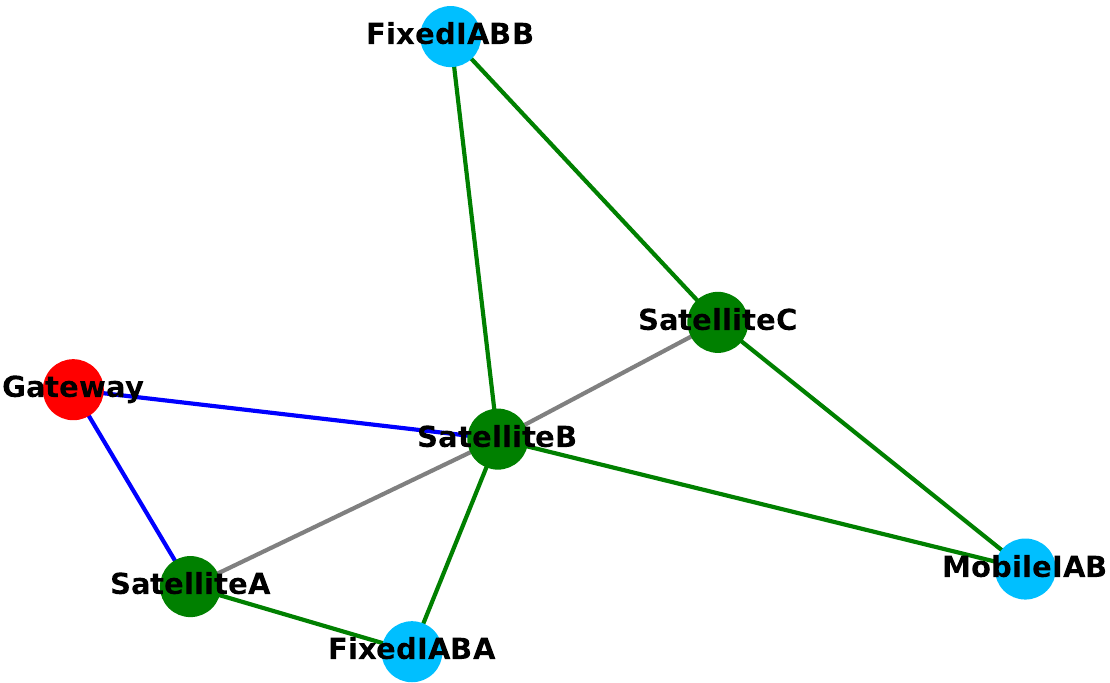}
		\caption{Interconnection between satellites and terrestrial IAB stations.} 
		\label{fig:Topology}
	\end{minipage}
\end{figure}
\begin{figure}[t]
	\centering
	\begin{minipage}{0.40\textwidth}
		\centering
		\includegraphics[width=1.0\columnwidth]{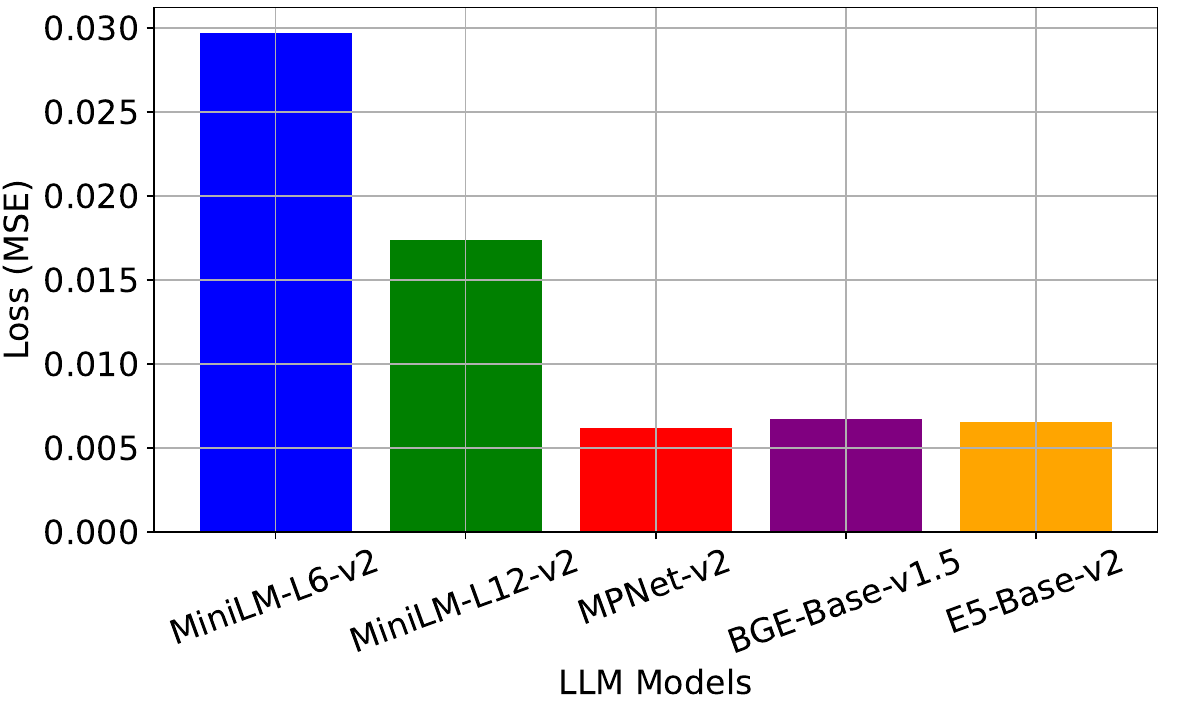}
		\caption{LLMs comparison in terms of MSE minimization.} 
		\label{fig:LLM_loss}
	\end{minipage}
	\begin{minipage}{0.40\textwidth}
		\centering
		\includegraphics[width=1.0\columnwidth]{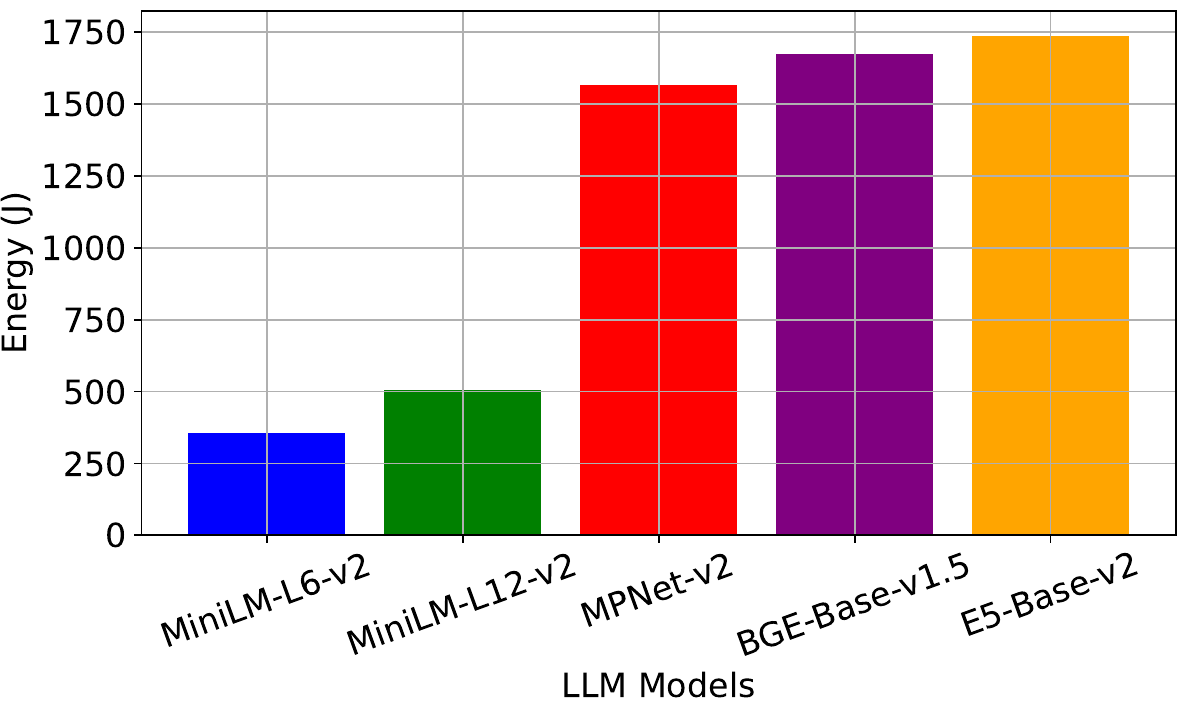}
		\caption{LLMs comparison in terms of energy consumption.}
		\label{fig:LLM_energy}
	\end{minipage}
\end{figure}
\begin{figure}[t]
	\centering
	\begin{minipage}{0.40\textwidth}
		\centering
		\includegraphics[width=1.0\columnwidth]{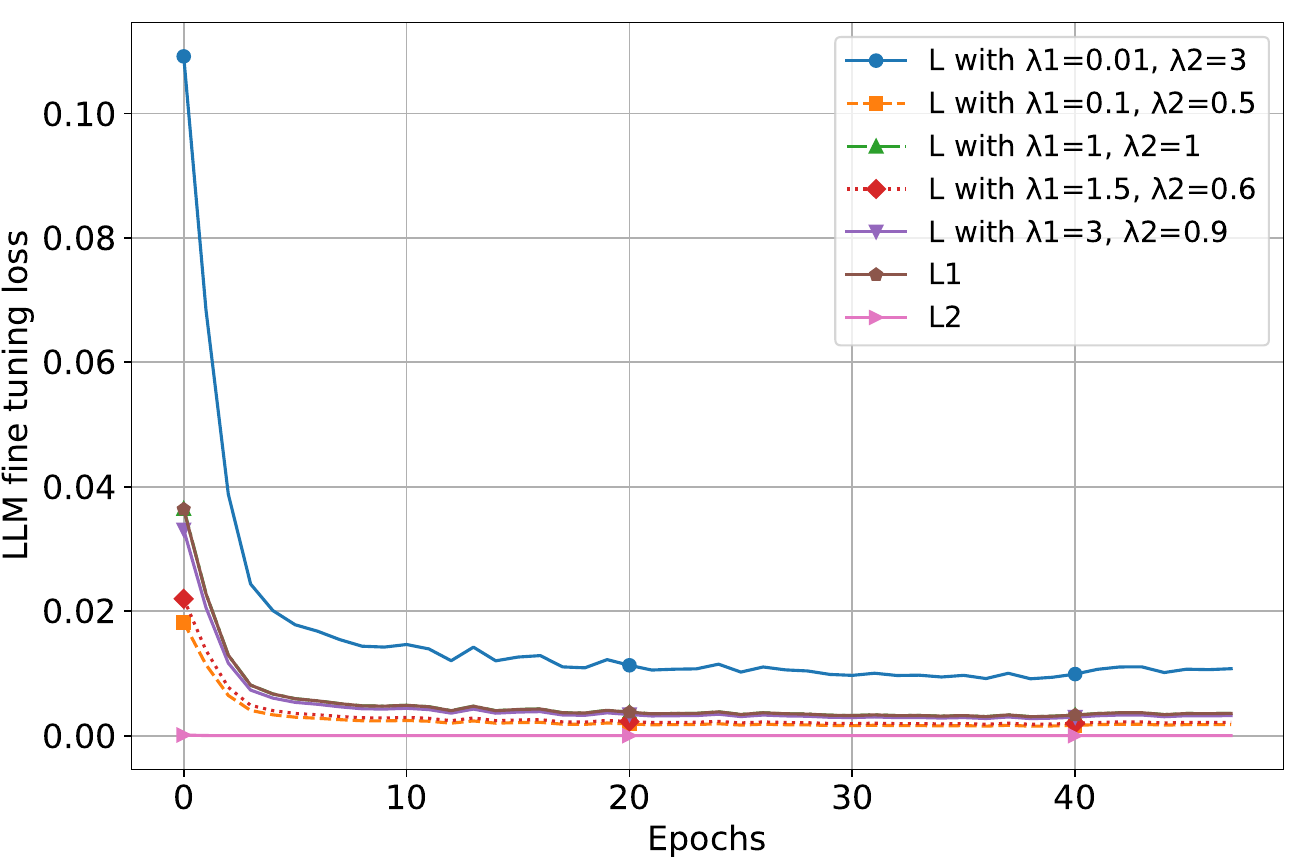}
		\caption{Loss function minimization for LLM fine tuning.} 
		\label{fig:LLM_fine_tuning}
	\end{minipage}
	\begin{minipage}{0.40\textwidth}
		\centering
		\includegraphics[width=1.0\columnwidth]{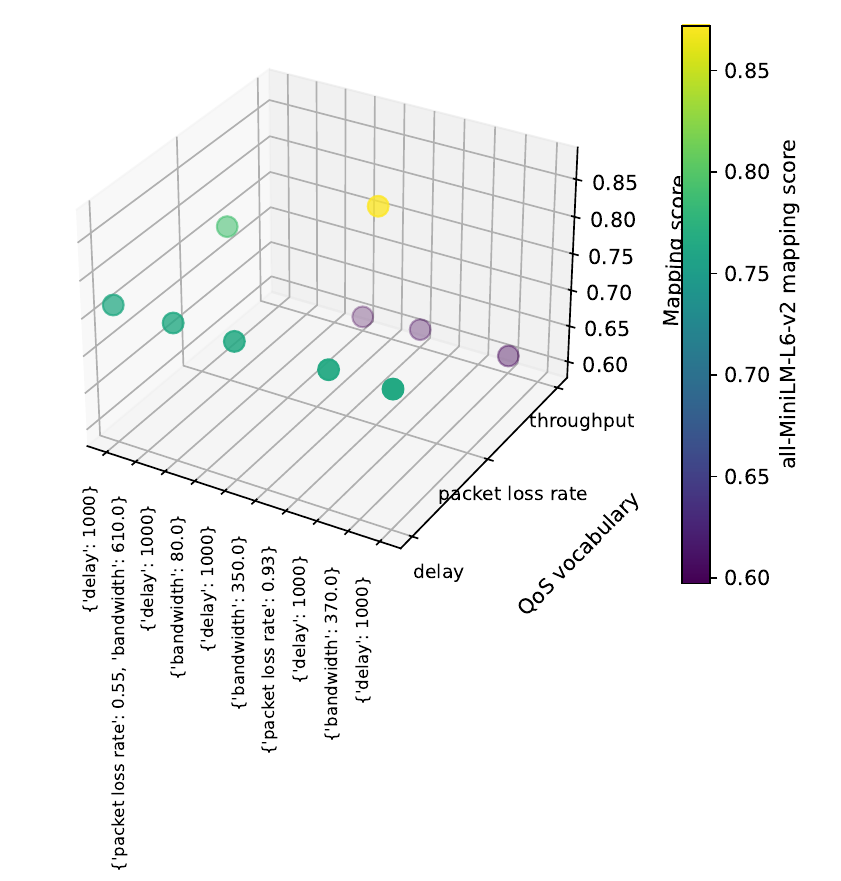}
		\caption{Mapping network vocabulary to QoS using LLM.}
		\label{fig:Network_to_QoS}
	\end{minipage}
\end{figure}
\begin{figure}[t]
	\centering
	\begin{minipage}{0.40\textwidth}
		\centering
		\includegraphics[width=1.0\columnwidth]{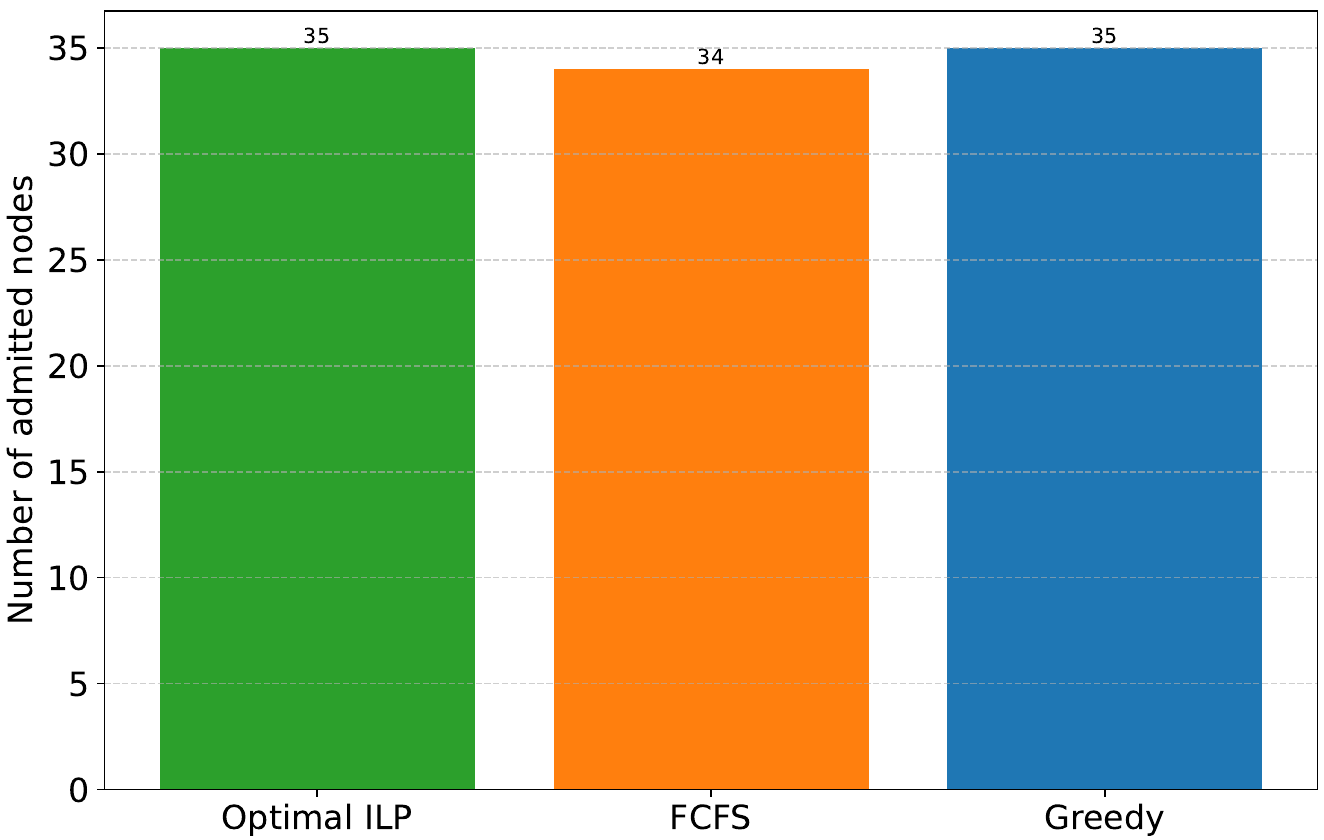}
		\caption{Summary of admitted nodes.} 
		\label{fig:AdmittedNode}
	\end{minipage}
	\begin{minipage}{0.40\textwidth}
		\centering
		\includegraphics[width=1.0\columnwidth]{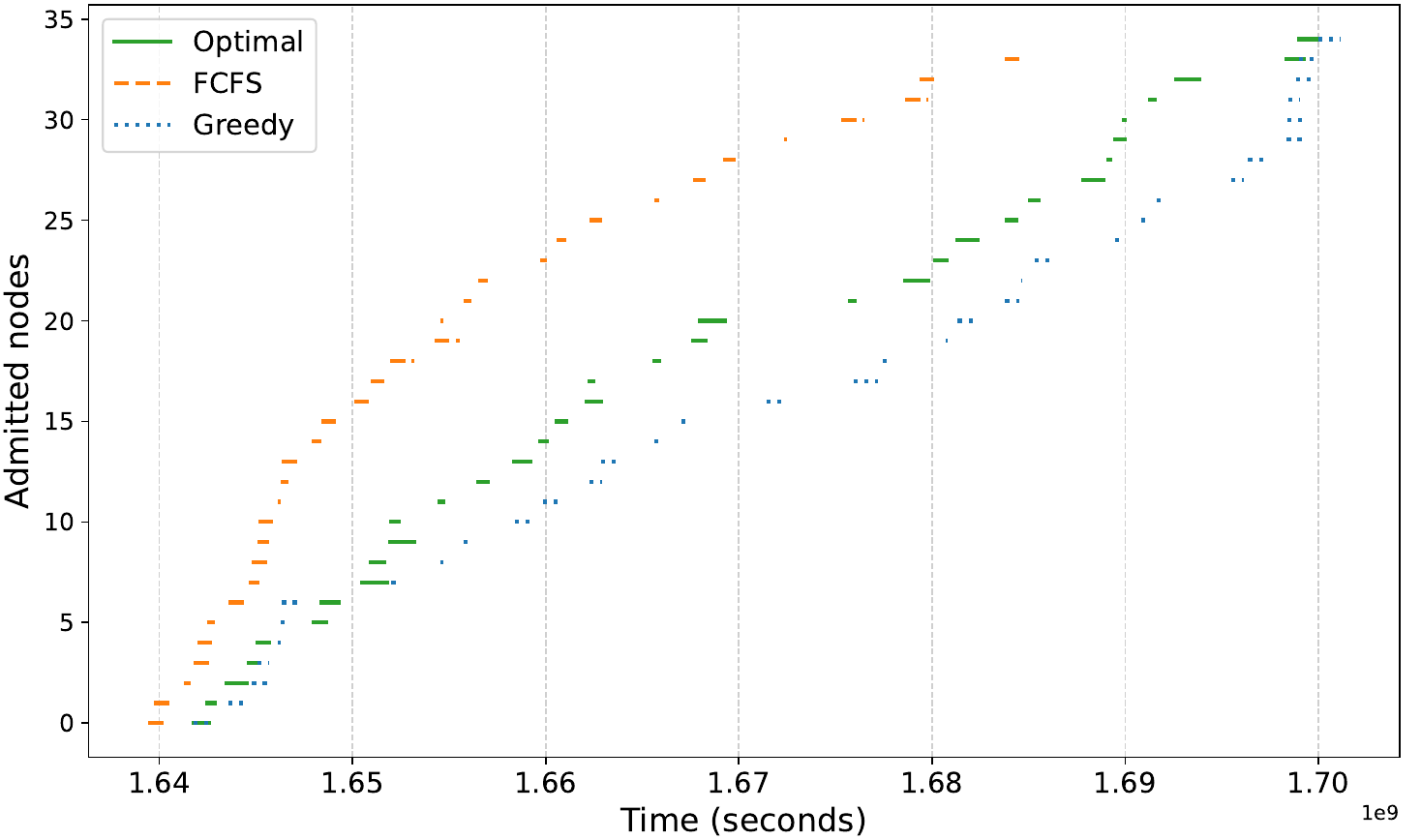}
		\caption{Admitted nodes over time.}
		\label{fig:AdmittedNodeOverTime}
	\end{minipage}
\end{figure}

\begin{figure}[t]
	\centering
	\begin{minipage}{0.40\textwidth}
		\centering
		\includegraphics[width=1.0\columnwidth]{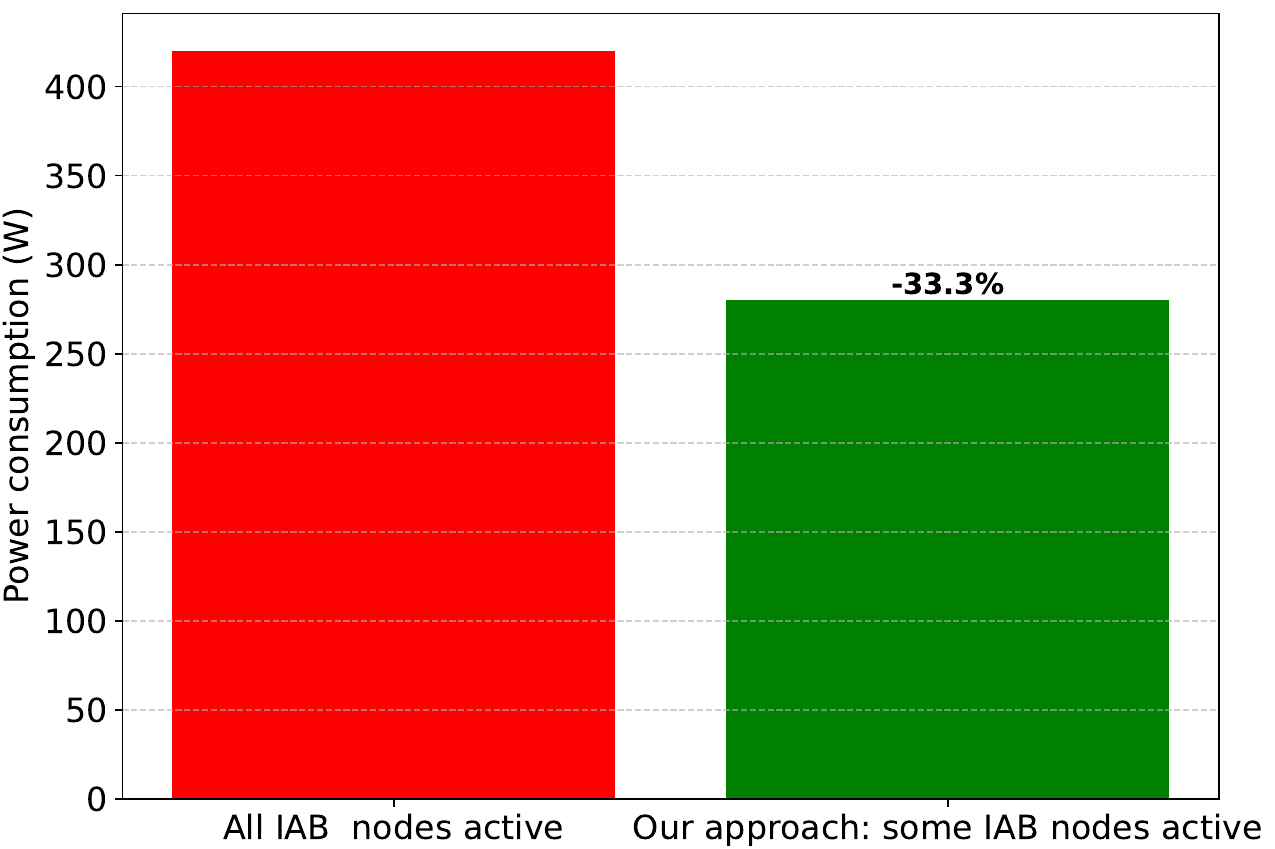}
		\caption{Power consumption based on IAB activation.} 
		\label{fig:PowerConsumption}
	\end{minipage}
	\begin{minipage}{0.5\textwidth}
		\centering
		\includegraphics[width=1.0\columnwidth]{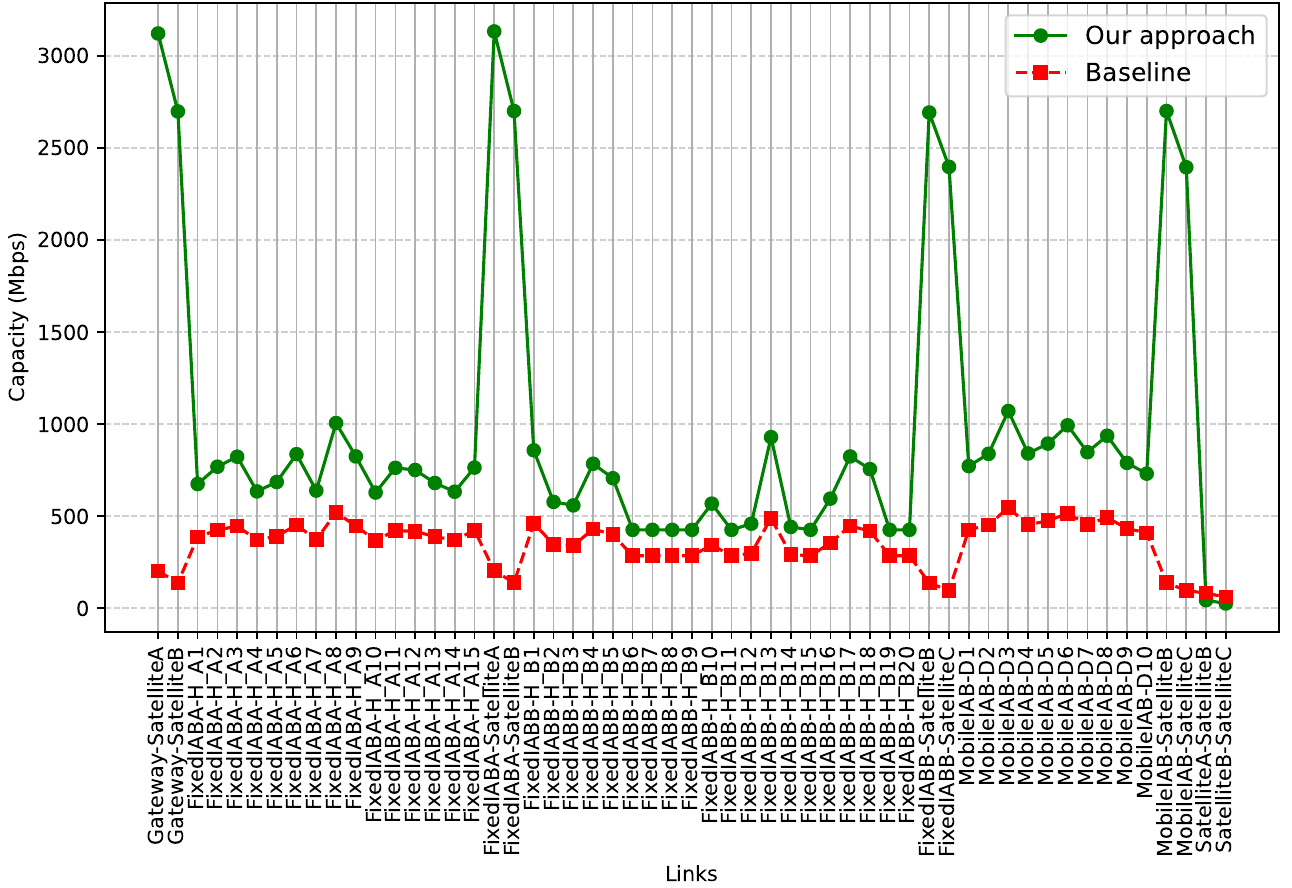}
		\caption{Data rate per link.} 
		\label{fig:Data_rate}
	\end{minipage}
\end{figure}

\begin{figure}[t]
	\centering
	\begin{minipage}{0.5\textwidth}
		\centering
		\includegraphics[width=1.0\columnwidth]{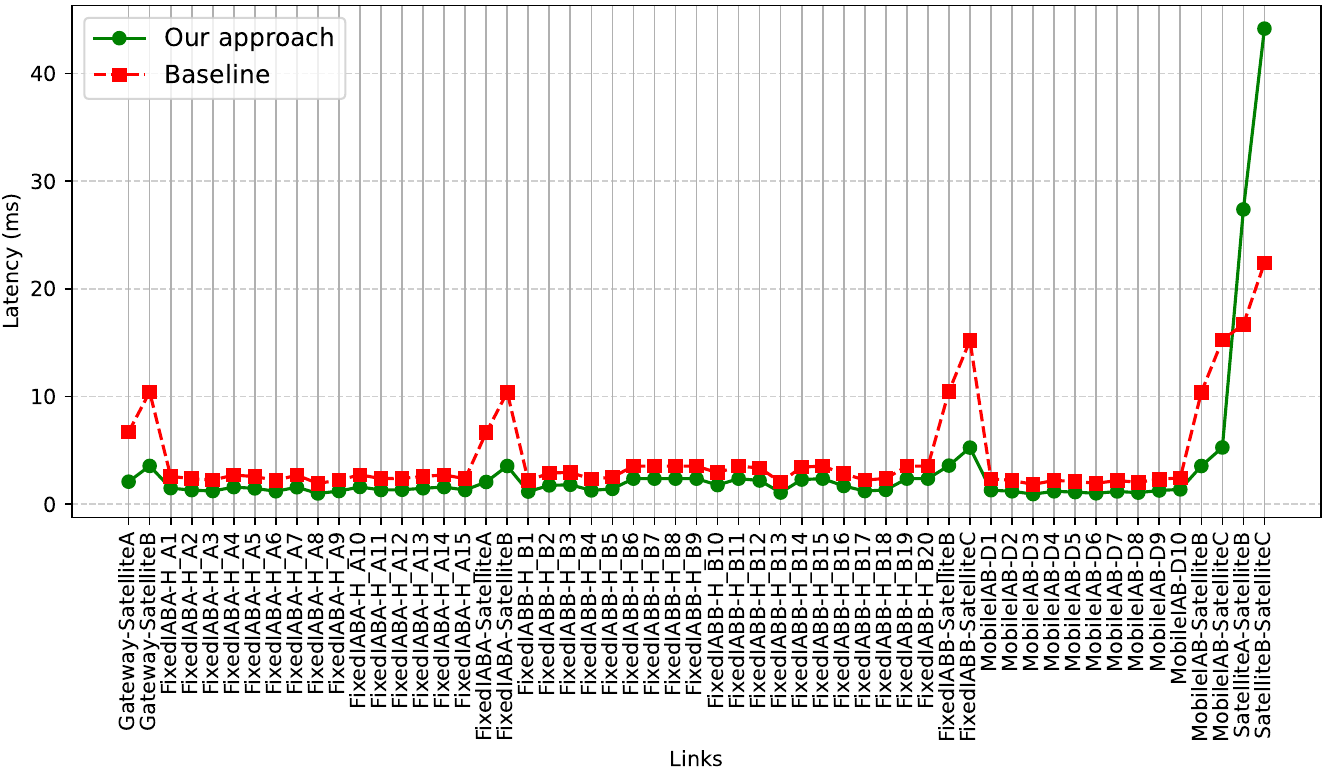}
		\caption{Latency per link.}
		\label{fig:Latency}
	\end{minipage}
	\begin{minipage}{0.40\textwidth}
		\centering
		\includegraphics[width=1.0\columnwidth]{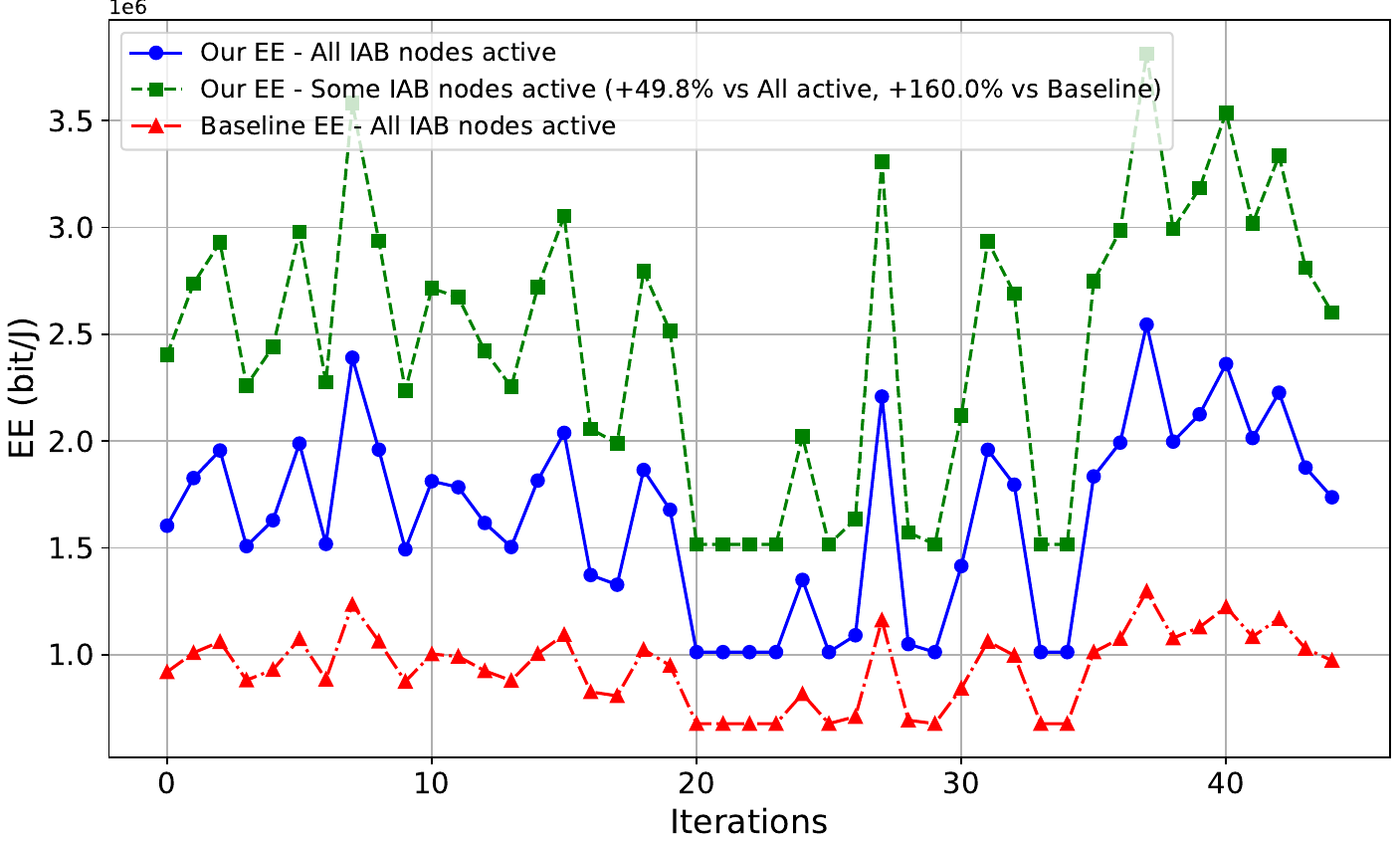}
		\caption{Energy efficiency based on IAB activation.}
		\label{fig:EnergyEfficiency}
	\end{minipage}
\end{figure}

In this section, we present the performance evaluation of
the proposed LLM-driven intent-aware Satellite–IAB–based FWA  approach to connect rural areas, where Python 3.10.12
language \cite{lutz2010programming} is used for numerical analysis. For intent-to-QoS mapping, we use all-MiniLM-L6-v2 from Hugging Face \cite{yin2024study, safikhani2025automl} as the sentence-transformer foundation LLM. For solving optimization problems, we use PuLP \cite{mitchell2009introduction}, a Python-based linear and mixed-integer programming modeler, with the GLPK solver \cite{gurski2017distributed}.
\subsection{Simulation Setup}
To create a network topology, we randomly choose a rural region in Kenya, Africa. The selected region is shown in Fig. 	\ref{fig:Region}.  In that region, we have two residential areas, each served by one terrestrial IAB node. Area A has 15 houses served by fixed IAB node A, while Area B has 20 houses served by fixed IAB node B.  Houses are in black, while IAB nodes A and B are in blue. The population of these residential areas uses a farming area located in the region, where IAB functionality is mounted on an agricultural truck as a mobile IAB to serve 10 connected Internet of Things (IoT) devices (these devices are in orange). In the region, we have one IAB donor in green. In other words, the IAB donor/gateway and fixed IAB node locations (latitudes and longitudes) are fixed.

To obtain the list of satellites and the times of each satellite pass in the region, we use the OneWeb satellite dataset from CelesTrak \cite{satellite} \cite{harvey2023tracking}, which includes 651 satellites. For loading the dataset and tracking satellites in the region, we use the Skyfield Python library \cite{Skyfield}. Based on the latitudes and longitudes of terrestrial fixed or mobile IAB stations, we continuously track and select three satellites that maximize the elevation angle $\theta^j$ to that region to establish intersatellite links connecting rural-area nodes to the core network via an IAB donor or gateway. Furthermore, each terrestrial IAB station selects two satellite links based on RSSI, with the strongest link being the strong link and the second link being the weak link. The interconnection between satellites and terrestrial IAB stations is shown in Fig. \ref{fig:Topology}.

The link capacities are computed based on the physical characteristics of each connection in the network graph. For every link, the geographical distance between nodes is first calculated using the Haversine formula. The distance is then used to estimate the signal propagation conditions. Each link is categorized into one of three types: laser intersatellite links, feeder or satellite-to-terrestrial backhaul links, and access links between fixed or mobile IAB nodes with CPEs or IoT devices. 
For instance, laser intersatellite links operate at 193 THz with 500 MHz bandwidth.
In contrast, feeder and satellite-to-terrestrial backhaul links use Ka-band (26.5 GHz) with 250 MHz bandwidth, while access links use sub-6 GHz (3.5 GHz) with 100 MHz bandwidth. The baseline scenario discussed in \cite{abdullah2023integrated, abdullah2024integrated} uses a 2 GHz carrier frequency and a 40 MHz bandwidth for inter-satellite, feeder, and satellite-to-terrestrial backhaul links.

To establish the temporal network in farming areas and to coordinate temporal and FWA in residential areas A and B, we assume that users in areas A and B express their interests as intents. For the intent expression, we use the intent dataset available in \cite{li2025business}. To map intent to the network vocabulary, we use the network vocabulary from \cite{kaggle} and the LLM all-MiniLM-L6-v2 \cite{yin2024study}. To map the extracted network vocabulary to network QoS, we use the 5G QoS requirements \cite{meredith2021management} and all-MiniLM-L6-v2. Therefore, among the many sentence-level LLMs available, the rationale for selecting all-MiniLM-L6-v2 over other models is discussed in the next subsection.

\subsection{Simulation Results}
For intent-to-network vocabulary mapping and mapping extracted network vocabulary to network QoS requirements, numerous sentence-level LLMs are available for semantic embedding tasks; selecting an appropriate model requires a systematic comparison of both semantic representation quality and computational efficiency. To this end, we evaluate five widely used sentence embedding models, namely all-MiniLM-L6-v2 \cite{yin2024study, wang2020minilm}, all-MiniLM-L12-v2 \cite{aperdannier2024systematic, wang2020minilm}, all-mpnet-base-v2 \cite{siino2024all}, BGE-Base-v1.5 \cite{bge_embedding}, and E5-Base-v2 \cite{wang2022text}. The comparison is conducted using semantic similarity learning. Specifically, each model is fine-tuned on sentence pairs with similarity labels, where semantically related pairs are assigned higher similarity scores and unrelated pairs are assigned lower scores. CosineSimilarityLoss is adopted during training to optimize the semantic embedding space. The evaluation considers both learning performance and computational cost. Learning performance is assessed using the mean squared error (MSE) between the predicted cosine similarities and the ground-truth similarity scores on benchmark sentence pairs. To assess computational efficiency, we measure training time, inference latency, CPU utilization, and estimated energy consumption. Energy consumption is approximated from CPU utilization and execution time using a CPU power model. Furthermore, a combined tradeoff metric is introduced by jointly considering normalized training loss and energy consumption, enabling the identification of models that achieve a favorable balance between semantic accuracy and computational efficiency. Fig. \ref{fig:LLM_loss} shows a comparison of LLMs in terms of MSE minimization, while Fig.\ref{fig:LLM_energy} shows a comparison of LLMs in terms of energy consumption. Since MSE is small across all LLMs, below 0.030, we choose LLMs that consume less energy, as the objective of this paper is to minimize energy consumption subject to performance constraints. Based on this comparison, the most suitable sentence-level LLM for the proposed semantic mapping task is all-MiniLM-L6-v2.

Using the intent and network vocabulary datasets, Fig.~ \ref{fig:LLM_fine_tuning} illustrates the fine-tuning loss function $L_1$ of the all-MiniLM-L6-v2 LLM model for mapping user intents to network vocabulary. In addition, the figure shows the $L_2$ loss function, which corresponds to the mapping from the extracted network vocabulary to network QoS. By jointly considering $L_1$ and $L_2$, the figure also depicts the composite loss function $L$ defined in~\eqref{eq:llmloss}. We chose $L$ with   $\lambda_1=0.1$,  $\lambda_2=0.5$ over other settings. After fine-tuning the all-MiniLM-L6-v2 model for the aforementioned tasks, Fig.~\ref{fig:Network_to_QoS} presents an example of network vocabularies extracted from user intents, mapped to corresponding network QoS requirements. Following this mapping, the required network QoS in rural areas is determined.

To provision such a network and estimate the number of users that can be served in rural areas under non-conflicting user intents, we solve the admission optimization problem defined in~\eqref{eq:conflict_corrected} using PuLP with the GLPK solver. We then compare the proposed solution with two baseline approaches: First-Come, First-Served (FCFS) and a greedy approach. In the greedy method, users are admitted based on their data rate requirements, starting with those having the lowest demand.
The results of this comparison are presented in Figs.~\ref{fig:AdmittedNode} and~\ref{fig:AdmittedNodeOverTime}. The results show that although FCFS is computationally efficient, it may reject users even when resources are still available, resulting in sub-optimal performance compared to the proposed optimal solution.

Fig.~\ref{fig:AdmittedNodeOverTime} shows that the number of terminals varies over time. During periods with fewer active users, certain IAB functionalities are deactivated to reduce energy consumption. By incorporating this adaptive activation mechanism, Fig.~\ref{fig:PowerConsumption} illustrates the resulting power consumption of the proposed approach, which dynamically deactivates IAB functions when the number of terminals is low, compared to a baseline scheme that keeps all IAB functions continuously active. The simulation results demonstrate that the proposed approach achieves up to a $33.3\%$ reduction in power consumption compared to the baseline.

Considering terminal connectivity across the network, Figs.~\ref{fig:Data_rate} and~\ref{fig:Latency} present the achievable data rate and latency, respectively. The results in these figures show that the proposed approach outperforms the baseline, achieving higher data rates and lower latency.
In both approaches, feeder (satellite-to-terrestrial) backhaul links exhibit higher latency compared to access links. This high latency is primarily due to the long propagation distance between terrestrial IAB nodes and satellite IAB nodes. Based on the achievable data rate and power consumption of IAB stations, Fig.~\ref{fig:EnergyEfficiency} compares the energy efficiency (EE) of the proposed approach with the baseline. The proposed approach considers deactivating IAB stations when the number of connected terminals is low, whereas the baseline assumes that all IAB functions remain continuously active.

\section{Conclusion}
\label{Conclusion}
This work proposes an LLM-driven, intent-aware satellite–Integrated Access and Backhaul framework for rural connectivity, designed to support both temporary field  and fixed household broadband access. By leveraging a large language model, the proposed framework translates users’ intents into explicit network requirements, enabling the network to adapt dynamically to highly heterogeneous, time-varying rural network demands. Based on inferred intents, we developed a satellite-IAB-enabled Fixed Wireless Access optimization framework that jointly manages node and link activation and traffic routing to improve energy efficiency while satisfying data-rate and latency requirements.
We formulated the problem as a mixed-integer linear programming (MILP) model and solved it using a two-stage Benders decomposition, thereby improving scalability for large-scale rural networks. Simulation results demonstrated that the proposed intent-aware framework significantly improves energy efficiency and resource utilization compared with conventional static rural networking approaches, while maintaining reliable connectivity for both household and field operations. The obtained results highlight the potential of combining LLM-based intent inference, satellite communications, and IAB technologies to enable adaptive, energy-efficient, and intelligent next-generation rural networks. Future work will investigate online learning mechanisms as replacements for traditional optimization approaches to reduce computational complexity, and will conduct experimental validation in realistic rural deployment scenarios.
\bibliographystyle{IEEEtran}

\end{document}